# GenDFIR: Advancing Cyber Incident Timeline Analysis Through Retrieval-Augmented Generation and Large Language Models

Fatma Yasmine LOUMACHI, Mohamed Chahine GHANEM *, and Mohamed Amine FERRAG

*Abstract*—Cyber timeline analysis or Forensic timeline analysis is critical in Digital Forensics and Incident Response (DFIR) investigations. It involves examining artefacts and events—particularly their timestamps and associated metadata—to detect anomalies, establish correlations, and reconstruct a detailed sequence of the incident. Traditional approaches rely on processing structured artefacts, such as logs and filesystem metadata, using multiple specialised tools for evidence identification, feature extraction, and timeline reconstruction. This paper introduces an innovative framework, GenDFIR, a context-specific approach powered by large language models (LLMs) capabilities. Specifically, it proposes the use of Llama 3.1 8B in zero-shot, selected for its ability to understand cyber threat nuances, integrated with a Retrieval-Augmented Generation (RAG) agent. Our approach comprises two main stages: (1) Data Preprocessing and Structuring: Incident events, represented as textual data, are transformed into a well-structured document, forming a comprehensive knowledge base of the incident. (2) Context Retrieval and Semantic Enrichment: A RAG agent retrieves relevant incident events from the knowledge base based on user prompts. The LLM processes the pertinent retrieved-context, enabling detailed interpretation and semantic enhancement. The proposed framework was tested on synthetic cyber incident events in a controlled environment, with results assessed using DFIR-tailored, context-specific metrics designed to evaluate the framework's performance, reliability, and robustness, supported by human evaluation to validate the accuracy and reliability of the outcomes. Our findings demonstrate the potential of LLMs in DFIR and the automation of the timeline analysis process. This approach highlights the power of GenAI, particularly LLMs, and opens new possibilities for advanced threat detection and incident reconstruction.

## I. INTRODUCTION

In recent years, a significant rise in cyber incidents has been driven by exposed vulnerabilities affecting a broad range of digital devices such as computers, IoT devices, network hardware (including routers, switches, and IDS), and embedded systems. After a cyber incident, a DFIR investigation is conducted to uncover the complexities of the attack [1]. This process begins with the collection of digital artefacts, followed by the extraction of reliable evidence, and concludes with identifying the incident's root cause. A crucial aspect of this investigation is timeline analysis, which focuses on examining the temporal sequences and chronological order of events. This involves identifying anomalies and suspicious patterns to reconstruct a comprehensive timeline of the entire incident [2].

The process of timeline analysis has traditionally been time-consuming due to the sheer volume and heterogeneity of collected data, requiring multiple specialised tools. Digital artefact analysis, as the first step in timeline analysis, is often performed using tools like Velociraptor [3], FTK [4], EnCase [5], Dissect [6], and others. These tools process structured data, such as logs and filesystem metadata, to extract relevant and reliable evidence from many events.

Following this, events are reconstructed to produce a coherent timeline that provides context and meaning to the incident. Specialised tools such as Timesketch [7] and Log2Timeline (Plaso) [8] are widely employed in this phase, as they facilitate the reconstruction and visualisation of timelines, simplifying the correlation of activities and enhancing the interpretability of events. As an example, Splunk [9] represents an advanced platform that offers features powered by AI and ML to assist in detecting anomalies, identifying unusual patterns in large datasets, and providing deep insights for incident investigations. Aside from tools that utilise AI for automation, other research methodologies and approaches to advance timeline analysis have been introduced [10], [11], [12], as evidenced in the literature.

Recently, generative AI models (GenAI), such as Large Language Models (LLMs), have emerged as a transformative force, surpassing traditional AI solutions. These models have been integrated and utilised across various fields, processes, and tasks and have also been proposed to automate aspects of digital forensics, incident response, and cybersecurity.

LLMs, including GPT [13], Llama [14], and Claude [15], excel in processing data, detecting anomalies, and generating natural language explanations, making them valuable for assisting with DFIR artefact analysis. However, their direct application to cyber incident timeline analysis remains unresolved. Furthermore, according to DFIR standard practices, traditional tools are still the preferred choice for managing structured artefacts such as logs, filesystem metadata, and binary data.

Despite their promise, LLMs face challenges such as hallucinations [16], memory limitations [17], and gaps in context-

* Mohamed Chahine Ghanem is the corresponding author.

F. Y. Loumachi is with Cyber Security Research Centre, London Metropolitan University, London, UK e-mail: f.loumachi@londonmet.ac.uk

M. C. Ghanem is with the Cybersecurity Institute, University of Liverpool & Cyber Security Research Centre, London Metropolitan University, UK e-mail: mcghanem@liverpool.ac.uk and m.ghanem@londonmet.ac.uk

M. A. Ferrag is with the Department of Computer Science, Guelma University, Algeria e-mail: ferrag.mohamedamine@univ-guelma.dz

specific knowledge. Innovations like Retrieval-Augmented Generation (RAG) address these limitations by integrating external knowledge bases, enhancing the contextual accuracy and reliability of LLM outputs [18]. Additionally, their effectiveness can be further improved through prompt engineering [19], especially in decoder-based models [20], and the deployment of task-specific agents.

### A. Research Question

As per the definition of timeline analysis and the tools and techniques present in the practices of DFIR, there is a gap where current methods face many limitations. For instance, existing solutions often present a final timeline without sufficient semantic context, where only timestamps of events are correlated. Furthermore, the use of multiple tools for analysis and others dedicated to event correlation complicates the process. Another challenge lies in the difficulty of spotting and successfully extracting evidence that could support claims regarding the incident and its root cause. While advancements in GenAI, particularly LLMs, and techniques like RAG, which enhance LLMs with external knowledge, offer promising potential to address these limitations, their application to DFIR timeline analysis remains underexplored. This study seeks to answer the following research questions:

- **RQ1:** How can LLMs and RAG be leveraged to enhance the current automation in Cyber Incident Timeline Analysis ?
- **RQ2:** How can a framework driven by RAG and LLMs advance incident timeline analysis by seamlessly integrating artefact analysis and event correlation ?
- **RQ3:** How can the framework be optimised to produce a reliable, comprehensive, and semantically-rich timeline for DFIR investigations ?
- **RQ4:** Is Timeline Analysis truly ready for Generative AI automation?

The paper is organised as follows to address the research questions: Section I introduces the field of DFIR and LLMs, establishing a foundational understanding of their core aspects. Section II expands on these details, offering deeper insights and a tailored definition relevant to this research. We then review related works and literature to build a robust foundation and gather additional perspectives. Section III describes the methodology of our framework, including its primary functions. Section IV discusses the implementation and testing of GenDFIR using synthetic scenarios. Section V presents the results and outputs generated by GenDFIR. Due to the size of the produced report, only one scenario is showcased in the paper. This section also evaluates the framework's reliability, functionality, and effectiveness to assess its performance. Section VI considers the limitations and ethical aspects of developing and deploying this framework, as well as its potential for expansion and adoption in real-world DFIR scenarios. Finally, Section VII concludes the paper with recommendations for refinement and suggestions for future research.

### B. Novelty and Contribution

In this paper, we propose a novel solution for interpreting cyber incident events in a human-understandable way, extending beyond traditional artefact analysis and conventional event correlation and timeline reconstruction. This is achieved through a DFIR-specific RAG agent powered by the Llama 3.1 8B model, designed to streamline cyber timeline analysis and address the limitations of current tools and techniques.

Our contribution lies in advancing the automation of critical tasks and processes within the field of DFIR, aiming to enhance the overall efficiency and effectiveness of cyber incident response and analysis.

## II. RESEARCH BACKGROUND AND RELATED WORKS

The proposed GenDFIR framework combines different technologies to leverage Generative AI in automating a DFIR task. To provide clarity on the application of RAG and LLMs in cyber incident timeline analysis, this section defines essential concepts and discusses related works. Some definitions have been adapted to suit the specific context of this research:

### A. Timeline Analysis in DFIR

**Digital Artefact:** There is no formal or precise definition of this term in the literature [21]. However, within the field of DFIR, artefacts are typically described as processed and relevant data collected and extracted from digital devices. For instance, in operating system (OS) forensics, these artefacts may include the file system, OS executables, network activity, internet history, cache, and other related data.

**Cyber Incident Event and Anomaly:** Refers to an action that potentially compromises or alters the security state of a system. In the context of a cyber incident, such actions are aimed at breaching the system's security policies [22]. An anomaly refers to deviations in behaviour that differ from established norms or expected patterns within a dataset. In the context of a cyber incident, an anomaly typically represents deviations from the expected normal behaviour of a system [23]. For instance, in Windows Event Logs, this could include unusual logon attempts, irregular application activity, or unexpected changes in system configurations.

**DFIR:** is a term that encompasses two essential processes: Digital Forensics (DF) and Incident Response (IR).

- *DF:* involves the management and analysis of digital evidence from its initial discovery to its presentation in legal contexts. This process includes the identification, collection, and analysis of evidence, with a key component being timeline analysis. Timeline analysis is essential in DF as it helps reconstruct the sequence of events by establishing the chronological order of actions, which is crucial for uncovering critical details and understanding the flow of the incident.
- *IR:* On the other hand, refers to the set of actions and procedures an organisation follows to detect, manage, and mitigate cyber incidents. It typically starts with preparation, followed by detection, where timeline analysis also plays a role in understanding the sequence of events and determining the scope of the incident. During the analysis phase, establishing a chronological order of actions helps assess how the incident progressed. The process continues with containment, eradication, recovery, and post-incident activities, where insights from timeline analysis can guide decisions and help evaluate the effectiveness of the response.

DF plays a crucial role in IR, providing in-depth technical analysis that supports the overall process. DF helps identify the root cause of an incident, detect hidden access points, and uncover malicious activities. While DF focuses primarily on the technical aspects of an incident, IR addresses the broader scope, including containment, vulnerability remediation, and managing both technical and organisational elements.

Together, DF and IR complement each other in addressing a cyber incident. For instance, while IR may struggle with handling legal and regulatory aspects, DF excels in this area by ensuring the proper collection and presentation of evidence. Thus, timeline analysis not only supports each process individually but also strengthens their collaboration in responding effectively to incidents [2].

**Timeline analysis:** Involves the presence of a range of sub-activities to construct a coherent timeline of a cyber incident. Many studies in the literature focus on individual tasks, such as artefact analysis, anomaly detection, event correlation, or timeline reconstruction in isolation, with some incorporating modern AI-based solutions. Unfortunately, few works address the integration of all components of timeline analysis, particularly with an emphasis on automation. Table I introduces notable works relevant to the field of full timeline analysis automation:

### B. Large Language Models

LLMs are advanced models of GenAI designed to understand and generate human language. They can predict word sequences and generate new text based on input data. Distinguished by their vast training datasets and sophisticated architectures, LLMs go beyond mimicking human-like creativity. They play a transformative role across various domains by enhancing productivity and automating complex tasks that would traditionally require human ingenuity [28]. However, a key distinction must be kept in mind, as various types of LLMs have been introduced, each optimised for specific tasks. For instance:

**Decoder-based models:** Such as GPT, LLaMA, Mistral, and Microsoft Phi excel in text generation, where the output is based on user input and the initial prompt:

- *Prompt Engineering:* is the art and science of skillfully crafting and designing prompts to maximise the capabilities of a model. In the context of timeline analysis, this involves strategically framing inputs, specifying technical details, and establishing the investigative context to refine the model's output based on its precise and meaningful definition [20]. For example, a DFIR analyst might prompt the system as follows: *"Conduct artefact analysis, correlate events, and reconstruct a coherent timeline of the incident."*. This method aims to ensure that the

LLM understands the context, adheres to specific DFIR timeline analysis constraints, and achieves the intended goals of the investigation and analysis.

These models use an autoregressive decoder architecture, which generates one token at a time, predicting the next token based on the previous ones:

- *Token:* In LLMs, a token represents a character, word, subword, symbol, or number [20]. However, in timeline analysis, a token can be represented as follows:
  *A non-tokenised Windows event log:*

```
Event ID: 7124, Details: Unusual network response pattern, Level:
Error, Date and Time: 08/01/2024 15:43:21, Source: Intrusion Detection
System, Task Category: Threat Detection.
```

Fig. 1: Windows Event Log Sample

*A tokenised form (using the text-embedding-ada-002 - external embedding model):*

Fig. 2: text-embedding-ada-002 - Windows Event Log Embedding (Hugging Face Tokeniser [29])

*A tokenised form (using the GPT3 - internal embedding model):*

Fig. 3: GPT3 embedding - Windows Event Log Embedding (Hugging Face Tokeniser [29])

  This process occurs during the tokenisation phase, which is part of the LLM architecture. The specific tokenisation and embedding approach used depends on whether the model relies on its internal system or integrates an external embedding model. Also, each model produces different outputs based on its tokenisation and embedding techniques [20].

- *Embedding:* is the process of converting text into numerical representations, often in the form of tensors, suitable for the LLM. It begins with tokenisation, where words or characters within a text are transformed into tokens, representing individual units. These tokens are then mapped to numerical values that capture their semantic representation. Subsequently, additional layer transformations and processing are applied to refine these representations further. The final output is a dense vector, where each value corresponds to a specific feature of the text [30].

This makes them particularly suited for generating coherent and contextually relevant text in response to user queries [31].

On the other hand:

**Encoder-based models:** For instance BERT, excel in tasks like classification and sentiment analysis, where they are commonly used in applications such as text categorisation and emotion detection. In these models, the output is generated based solely on the information present in the input, without relying on external context or previous outputs .

| Finding | Approach | Overview |
|---------|----------|----------|
| **Tool:** Eric Zimmerman's tools [24] | Processing various types of data, including event logs, registry entries, and metadata, to provide detailed insights into incidents. | Beyond the tools discussed earlier, others, [24], have gained recognition for their capabilities in performing timeline analysis at a deterministic forensic level. However, they are not AI-based and lack automation, relying heavily on the expertise of the analyst or investigator. |
| **Study:** Chabot et al. 2014 [25] | Approach begins with data collection from diverse sources, whether online or from devices. Raw data is examined using tools such as Zeitline, which analyses digital artefacts, and log2timeline, which automates the extraction of events from these artefacts. Subsequently, the extracted events are managed using the FORE system, which addresses semantic heterogeneity and stores events in their natural language form. Semantic processing transforms the raw data into knowledge using ontology-based tools. Custom algorithms then correlate events by identifying and computing common subjects and objects while analysing their temporal proximity to establish relationships and patterns. Finally, the framework employs graphical representations to visualise the sequence of events and their correlations, enhancing interpretability. | This contribution proposes a systematic, multilayered framework focusing on semantic enrichment to tackle challenges in timeline analysis. This approach not only automates timeline analysis but also delivers semantically enriched representations of incident events. However, one apparent limitation is the reliance on multiple standalone tools, which may complicate the workflow. |
| **Study:** Bhandari et al. 2020 [26] | Techniques that primarily involve managing, organising, and structuring temporal artefacts into a more comprehensible timeline. Log2timeline is utilised to extract timestamps from disk image files, while Psort processes the output to further handle the temporal artefacts and generate the final timeline. | Authors introduced an approach that addresses the complexities and challenges of understanding generated temporal artefacts using abstraction techniques. The foundation of this research lies in structuring data for better management and faster timeline reconstruction. It is important to note that in this approach, artefact analysis is performed manually. Although the methodology successfully manages the textual nature of events and produces easily interpretable results, it still relies on manual intervention for analysis. |
| **Study:** Christopher et al. 2012 [27] | Achieved by proposing the use of analyser plugins to conduct detailed analysis on raw, low-level events. These plugins extracted relevant data and aggregated it into high-level events. They then used Bayesian Networks to correlate and link these high-level events by performing probabilistic inference. | The study focuses on automating event reconstruction and generating a human-understandable timeline. This was The main advantage of this approach is its ability to successfully handle and process large volumes of data, as well as produce an interpretable timeline. |

TABLE I: DFIR Timeline Analysis

Additionally, there are:

**Encoder-decoder models:** Like T5 and BART, which excel in tasks such as text translation and summarisation, as well as speech recognition and image recognition. The output of such models is based on both the input and the context, where the output is a transformed version of the input [32].

In the field of cybersecurity and DFIR, several studies have been conducted, and approaches have been proposed that integrate LLMs for various applications as illustrated in Table II:

Based on available findings, the most widely used state-of-the-art models in the scope of this research are GPT and Llama. In study [42], researchers disclosed the capabilities of Llama 3.1 in performing advanced cybersecurity tasks, trained on publicly validated data, including cybersecurity-related content. An important consideration is the benchmarking over the new CYBERSECEVAL 3 suite to measure cybersecurity risks and capabilities. The model can recognise and identify cyber threats with high accuracy.

In addition, GPT has been used in different studies, as shown in TABLE II , for the purpose of identifying and mitigating cyber threats. Furthermore, in study [43], the authors mentioned its ability to perform even in a zero-shot setting, as it has been extensively trained on massive datasets, such as reports, journal and conference papers, previous DFIR cases, standards, frameworks, guidelines, and conversations related to the field. Examples include IEEE articles, Wikipedia entries, CTDD, IDS2017 and IDS2018 datasets, ADFA (Australian Defence Force Academy) datasets, OpenTitan System-on-Chip (SoC) data, Hack@dac 2021 SoC data, CVE reports, and more [33].

### C. Retrieval-Augmented Generation (RAG) in Information Extraction and Knowledge Expansion

**RAG:** Is a technique that primarily aims to to optimise and enhance the performance of LLMs. It utilises an external knowledge base beyond the LLM's pre-existing knowledge and training datasets to provide additional information during inference. This process is not the same as fine-tuning or training an LLM, but rather involves dynamically retrieving, generating, and integrating relevant external textual information from databases (external storage), vault files (internal storage), or through cloud pipelines. It is crucial to acknowledge that while an LLM alone is very powerful, its knowledge may not encompass specific contexts [44] [18].

RAG consists of two essential functions, retrieval and augmented generation:

- *Retrieval-Augmented:* In the context of RAG, retrieval refers to the process of searching and selecting relevant information from a knowledge base or document dataset to enhance the output of a generative model. This ensures the model generates more accurate and contextually relevant responses [44]. Retrieval techniques commonly used in information retrieval include cosine similarity,

| Finding | Approach | Overview |
|---------|----------|----------|
| **Study:** Ferrag et al. 2024 [33] | Reviewing and examining research studies, published articles, and journals addressing the integration of generative AI and LLMs into cybersecurity. Additionally, the paper discusses features, findings, insights, and theoretical approaches derived from datasets relevant to cybersecurity, including training methods, architectures, and associated mathematical equations. | The authors provide an extensive review of the application of LLMs in the field of cybersecurity, covering its subfields, including intrusion detection, cyber forensics, and malware detection. |
| **Study:** Otal et al. 2024 [34], Wang et al. 2024 [35], Fariha et al. 2024 [36], Saha et al. 2024 [37] | - [34] integrates LLMs, employing various decoder-based models such as LLaMA3 8B and 70B, and Phi3. Their findings showed that smaller models, like the 8B, proved efficient when applied to honeypot systems for conducting advanced malicious activity analysis and detection <br> - [35] proposed combining LLMs, specifically GPT and BERT, with LSTM to predict cyberattacks in IoT networks. <br> - [36] explored the use of GPT-3.5 for log summarisation to analyse and summarise log files and detect specific events. <br> - [37] introduced an advanced paradigm utilising GPT for SOC tasks, including vulnerability insertion, security assessment, and security verification. | All of these studies explored the use of different LLMs to automate specific cybersecurity tasks by embedding them into their workflows. |
| **Study:** Wickramasekara et al. 2024 [38] | Theoretically introducing and explaining how LLMs can be utilised in various phases of a DF investigation and how specific models can perform particular tasks. For example, a model like GPT-3.5 can generate textual reports at the conclusion of investigations, while multimodal LLMs, such as GPT-4 and LLaVA with vision assistance, can analyse images and videos, providing contextual outputs for digital forensics. | The paper provides an extensive literature review on the integration of LLMs to advance the DF process. |
| **Study:** Scanlon et al. 2023 [39] | The role of ChatGPT in supporting various tasks, including artefact analysis, generating regular expressions and keyword lists, creating scripts for file carving, RAID disk acquisition, and password cracking, identifying IR sources, anomaly detection, and developing detailed forensic scenarios. | The article presents a comprehensive study on how ChatGPT can assist during DF investigations, examining this concept from multiple perspectives. The paper also addresses the limitations and strengths of ChatGPT, clearly stating that while the model significantly enhances the DF process, its current stage of technological development still leaves it vulnerable to biases and its non-deterministic nature. As a result, human oversight remains essential. |
| **Study:** K et al. 2023 [40] | ChatGPT, powered by GPT-4 and GPT-3.5 models, to analyse artefacts (input data) and extract relevant evidence, such as conversations, images, and other information pertinent to the investigation. | The paper proposes using ChatGPT to enhance digital investigations by identifying evidence during the DF process. The paper emphasises that, despite the efficiency of this method, its outputs must always be verified and monitored by humans. |
| **Tool:** CADO 2024 [41] | Dedicated to assisting forensic analysts with investigations by providing insights and streamlining the investigation process. | A recent AI-based platform CADO [41] powered by a local LLM has been developed. |

TABLE II: Large Language Models

Jaccard similarity, BM25, TF-IDF, Latent Semantic Analysis (LSA), and embedding-based models. These methods evaluate the similarity between the query and documents, retrieving the top k pieces with the highest similarity scores to provide relevant contextual knowledge for the LLM [45].

In GenDFIR, retrieval is applied to extract relevant incident events from documents containing cyber incident events. It helps identify and select the most pertinent ones (Top k pieces) based on their relevance to the query or event context.

- *Augmented Generation:* Is the process where, after retrieval, the extracted information is provided to the LLM for semantic enhancement, enabling it to generate more accurate and contextually enriched outputs [44]. In the case of GenDFIR, this involves extracting evidence (mostly anomalous events) from the incident knowledge base and passing it to the LLM to produce a timeline of the incident, which includes key events, correlations, and analysis of the evidence.

**Knowledge Base:** In LLMs with RAG, a knowledge base refers to a structured repository of factual knowledge and data related to specific domains, which the model may use during inference [18]. In the context of this research (GenDFIR), it will serve as the repository for storing and managing all data generated and collected from a cyber incident. This data is presented in a natural language format, with all data stored in English.

**Top-k Pieces/Evidence:** Top-k is a hyperparameter in RAG that controls the number of the most relevant pieces of information retrieved from an external knowledge base during the retrieval phase. The value of k determines how many chunks or documents, ranked by relevance, are returned for further processing by the LLM [46].

In GenDFIR, the top-k pieces represent the most relevant retrieved log events (or evidence) with the highest anomaly scores, pertinent to the cyber incident. Initially, k refers to all events within the knowledge base, and the top-k represents the most relevant ones selected for further analysis by the LLM.

**Chunk:** A chunk is a manageable segment of text extracted from a document for specific processing or analysis. Chunking involves dividing large texts into smaller segments that can be embedded through an embedding model. In timeline analysis, chunking ensures that critical details, such as timestamps and numerical tokens representing seconds or minutes, are accurately captured. These small details are vital for event correlation, where even slight variations in timing can significantly impact the analysis. By segmenting the text into manageable chunks, these details can be preserved and embedded effectively, enabling precise token calculation and retrieval from an external knowledge base when using RAG with LLMs. The precision of timeline analysis depends on the segmentation method used. Factors like token calculation, and the careful selection of the embedding model all influence the accuracy and reliability of the process. Several reliable approaches to chunking have been proven effective [52]:

- *Token-Based:* Divides text based on a fixed number of tokens.
- *Paragraph-Based:* Segments text by paragraphs to maintain context.
- *Semantic-Based:* Groups text based on meaning or topics.
- *Sentence-Based:* Segments text into sentences, each of which may have unique semantics.

| Finding | Approach | Overview |
|---------|----------|----------|
| **Study:** Tihanyi et al. 2024 [47] | RAG to generate high-quality, context-based questions using an external knowledge. | Authors present in their paper how RAG is used to advance the process of creating a cybersecurity-based dataset. Subsequently, CyberMetric was used to benchmark the general cybersecurity knowledge of cybersecurity-oriented LLMs. |
| **Study:** Lála et al. 2023 [48] | Employing RAG agents to answer questions based on embedded scientific literature. | Paper explains and showcases the power of RAG agents in addressing some limitations of traditional LLMs, such as hallucinations and lack of interpretability. |
| **Study:** Wang et al. 2023 [49] | RAG extraction and retrieval are optimised using chunks and tokens, where chunks are used for text segmentation, and tokens represent units, words, or subwords within a chunk. | Authors propose an approach to enhance the performance of RAG in single LLMs for both contextual ranking and answer generation. The goal is to ensure that each chunk contains sufficient context to answer questions and queries accurately. |
| **System:** Chat2Data by Zhao et al. 2024 [50] | A prototype for advanced data analysis using RAG for data retrieval, and a knowledge base where all data are stored. Outputs are shown in a graphical representation. | The authors applied RAG and LLMs to build and introduce an interactive system. |
| **System:** BIDARA by Toukmaji et al. 2024 [51] | Employing RAG technologies and LLM agents to address the complexities of biomimicry. | An AI research assistant model was presented as a system. |

TABLE III: Retrieval-Augmented Generation

**DFIR Context-Specific RAG Agent:** A RAG agent is a key component of some modern LLMs-powered AI systems. It acts as an intermediary between the LLM and the user, managing the search process when a query or input is received. The agent retrieves relevant information from a knowledge base by employing retrieval methods that measure data similarity. This context-specific information is then provided to the LLM, which uses it to generate a response tailored to the query. Additionally, as part of the process, the agent can augment generation by refining and modifying human-crafted prompts to enhance the quality of the generated response in line with the specific task. During inference, the agent processes the input, adjusts it based on its designated role, and helps the LLM produce a more relevant and accurate answer [53]. It is important to note that the role of a RAG agent is defined by a human, ensuring that the task aligns with its intended purpose. In the case of timeline analysis, an example could be summarised as: a DFIR analyst prompts the system with *"Conduct timeline analysis,"* while the agent's prompt (task) would be: *"You are a DFIR AI assistant, tasked with analysing artefacts, correlating events, and producing a coherent timeline of the incident. Base your answer on the provided context and do not include additional information*

*outside of the context given."*. This allows the RAG agent to extract relevant incident information, such as pertinent event logs, from the knowledge base. Additionally, the system can support the DFIR analyst by refining the prompt if critical details are overlooked, aiming for more optimised results.

Table III highlights several studies that have explored the use of RAG for information extraction and knowledge augmentation:

## III. RESEARCH METHODOLOGY

GenDFIR is an LLM-powered framework specifically proposed for timeline analysis in DFIR. However, using an LLM alone presents several challenges, particularly with respect to its context window limitations. LLMs, such as GPT, process input as large text blocks, which restricts the amount of detailed event data they can accommodate at once [54] [55]. This creates a significant issue in DFIR timeline analysis, where precise event-specific details are crucial. When processing DFIR incident documents, the LLM may overlook vital information, as its context window prioritises broader context over the granular details necessary for an accurate timeline reconstruction.

Furthermore, capturing the temporal and logical relationships between events in a cyber incident is a complex task, often overlooked in traditional methods. These approaches generally require repeatedly inputting artefacts into the LLM, each time referencing previous data. This process is not only labour-intensive but also inefficient, as the model must reprocess prior references without maintaining a coherent understanding of the evolving context of the incident.

To address these limitations, we propose integrating RAG technology within GenDFIR. Rather than enhancing the LLM with an external knowledge base, RAG will treat the DFIR incident reports themselves as the source of knowledge. These reports, containing detailed event data, will be directly uploaded into the framework, with the knowledge representing the interconnected events within the incident. RAG enhances the LLM's ability to correlate events accurately and efficiently, enabling dynamic extraction of events and providing easy, instant access to the incident data.

In addition to RAG, a RAG agent powered by the LLM will further augment the process by acting as a DFIR timeline analysis context-specific agent. This agent will manage the process of filtering and extracting information from the knowledge base based on its role. It will enable advanced event filtering, ensuring that only the most relevant events are retrieved for analysis in the specific context of the incident.

This approach not only overcomes the limitations of the context window but also surpasses traditional methods by providing interpretable and enriched contextual outputs, ensuring a more comprehensive and efficient timeline analysis.

The following sections further explain and advance the technical methodology of GenDFIR, outlining its development:

### A. Incident Events Preprocessing and Structuring:

After a cyber incident, various artefacts are generated and stored in CSV files, forming incident datasets. The preprocessing step involves converting these CSV files into a structured text document.

A single Event $E$ is is characterised and defined by its attributes $e$ as follows:

$$E = \langle e_1, e_2, \ldots, e_n \rangle \quad \text{where } n \text{ is the last number of attributes}$$

and $e_1$, $e_2$, $e_n$ are the individual attributes of the event, such as Level, Date and Time, Source, Event ID and Task Category.

We define a Cyber Incident $I$ as a set or sequence of Events $E$:

$$I = \langle E_1, E_2, \ldots, E_m \rangle \quad \text{where } m \text{ is the last number of Events}$$

A cyber incident $I$, represented by a CSV file contains $t$ events.

The CSV file is converted into a textual incident document $ID$, preserving the same events, and is structured as follows:

$$I_{\text{CSV}} = \langle E_1, E_2, \ldots, E_t \rangle$$

$$ID = \langle E_1, E_2, \ldots, E_t \rangle \quad \text{where} \quad ID = I_{\text{CSV}}$$

- The collected events refer to those that occurred during the timeframe of the cyber incident.

Each row of an event in $I_{\text{CSV}}$ is treated as an individual context, representing a unique event within the timeline of the incident. These rows are then converted into corresponding segments in $ID$ maintaining their sequential order and the integrity of their context throughout the transformation process.

The following figures illustrate the transformation of $I_{\text{CSV}}$ to $ID$:

*Unprocessed and Unstructured Windows Event Logs stored in CSV format:*

| | A | B | C | D | E | F | G |
|---|---|---|---|---|---|---|---|
| 1 | Event ID | Details | Level | Date and Time | Source | Task Category | |
| 2 | 7124 | Unusual network response pa | Error | 08/01/2024 15:43 | Intrusion Detecti | Threat Detection | |
| 3 | 8921 | Anomalous IP returned in DNS | Warning | 08/01/2024 13:12 | DNS Server | Query Response | |
| 4 | 5429 | Network packet inconsistency | Critical | 08/01/2024 16:55 | Network Firewall | Suspicious Traffic | |
| 5 | 3302 | Multiple DNS queries with del | Warning | 08/01/2024 10:24 | Intrusion Detecti | Network Performance | |
| 6 | 1203 | Outbound traffic to unexpecte | Error | 08/01/2024 18:32 | DNS Server | Unusual Traffic | |
| 7 | 6708 | Increased latency in DNS req. | Warning | 08/01/2024 11:46 | Web Proxy | Performance Degradation | |
| 8 | | | | | | | |

Fig. 4: Windows Event Log CSV

*Processed and Structered Windows Event Logs stored in a document (PDF) format:*

Event ID: 7124, Details: Unusual network response pattern, Level: Error, Date and Time: 08/01/2024 15:43:21, Source: Intrusion Detection System, Task Category: Threat Detection.

Event ID: 8921, Details: Anomalous IP returned in DNS query, Level: Warning, Date and Time: 08/01/2024 13:12:05, Source: DNS Server, Task Category: Query Response.

Event ID: 5429, Details: Network packet inconsistency, Level: Critical, Date and Time: 08/01/2024 16:55:40, Source: Network Firewall, Task Category: Suspicious Traffic.

Event ID: 3302, Details: Multiple DNS queries with delayed responses, Level: Warning, Date and Time: 08/01/2024 10:24:11, Source: Intrusion Detection System, Task Category: Network Performance.

Event ID: 1203, Details: Outbound traffic to unexpected server, Level: Error, Date and Time: 08/01/2024 18:32:45, Source: DNS Server, Task Category: Unusual Traffic.

Event ID: 6708, Details: Increased latency in DNS requests, Level: Warning, Date and Time: 08/01/2024 11:46:22, Source: Web Proxy, Task Category: Performance Degradation.

Fig. 5: Windows Event Log Document (Knowledge Base)

The format in which an event is written in the document plays a crucial role in enhancing retrieval precision and optimising the LLM's understanding.

In GenDFIR, we propose a dynamic structure based on the context of the incident. For instance, in the case of collected event logs from a Windows incident, the structure is as follows: each attribute of an event is separated by a comma (","), and the event concludes with a full stop (".") to signify that each event is separate and represents unstructured data. However, when the format of an event differs from that of Windows event logs (e.g., Linux syslog events, web server logs, firewall logs, or database logs), alternative structural models can be adapted based on the type and nature of the data. In incidents requiring more contextual analysis, such as phishing email incidents, each email is represented as a single event, with a symbol (e.g., "/") used to indicate the end of each email.

### B. Incident Events Chunking

Building on the previously defined concepts of chunking and the structured document containing the incident, we propose a DFIR context-specific chunking method designed to improve retrieval precision, with each event representing a distinct chunk:

1) **Cyber Incident Event Length:** The length of an event is defined as the total sum of the number of characters across its attributes. The characters in a line of events are counted based on their attributes, $e_{ij}$, where:

   - $i$ represents the event number (e.g., Event 1, Event 2), and
   - $j$ represents the attribute number within that event (e.g., Event ID, Level).

   The total number of characters in the $i$-th event is given as follows:

$$T(E_i) = \sum_{j=1}^{n} T(e_{ij}) \tag{1}$$

   - $T(E_{ij})$ is the total number of characters in the $i$-th event.

- $T(e_{ij})$ is the number of characters in the $j$-th attribute of the $i$-th event.
- $n$ is the total number of attributes in the $i$-th event.

2) **Events Chunking:** The proposed method extends the general chunking concept by incorporating the calculated lengths of individual events. It enhances retrieval precision through precise semantic embedding, as every event attribute contributes to the chunking process, improving the accuracy of retrieval tasks. The chunking process adheres to the following equations to ensure compliance with both the token capacity of the LLM and the adopted embedding model:

To calculate the number of tokens for an event or chunk, we consider the token capacity of the embedding model, for instance, models that can process up to 512 tokens per input.

For optimisation, we assume an average of 4 characters per token, which is within the standard range of 4 to 6 characters in English, with 4 selected to maximise token capture and improve input accuracy. Based on this, the number of tokens for an event $T_{tokens}(E_i)$ is approximately the total number of characters in the event $T(E_i)$, divided by the average of $C_{avg}$ characters per token. This relationship is expressed by the Equation 2:

$$T_{tokens}(E_i) \approx \frac{T(E_i)}{C_{avg}} \quad \text{where} \quad C_{avg} = 4. \quad (2)$$

- $i$ represents the index of the $i$-th event in the sequence.

Similarly, the number of tokens for a chunk $C_k$ is calculated in the same manner, as in Equation 3:

$$T_{tokens}(C_k) \approx \frac{T(C_k)}{C_{avg}}. \quad (3)$$

- $k$ represents the index of the $k$-th chunk in the sequence.

This equation follows the same principle as equation 2 but applies to a chunk $C_k$ (a chunk $C_k$ represents a single event $E_i$). $T(C_k)$ is the total number of characters in a chunk.

The constraint on the number of tokens in a chunk is introduced in Equation 4. It ensures that the number of tokens for any chunk $T_{tokens}(C_k)$ does not exceed the model's token capacity $TM$:

- $TM$ represents the model's maximum token capacity, such as 512 tokens.

This helps ensure that a chunk fits within the embedding model's processing limits.

$$T_{tokens}(C_k) \leq TM, \quad (4)$$

$$M_{max} \approx TM \times C_{avg}, \quad (5)$$

- $M_{max}$ represents the maximum number of characters that the model can embed per chunk.

We define a cyber incident $ID$ in a document as the sum of all chunks $C_i$, where each chunk corresponds to an event $E_i$:

$$\text{ID} = \begin{cases} \sum_{i=1}^{n} C_i, \\ \text{where} \quad C_i = E_i, \\ \text{and} \quad E_i = \sum_{j=1}^{m_i} \text{Character}_{ij} \\ \text{with} \quad m_i \leq Mmax \end{cases} \quad (6)$$

- $n$ is the total number of chunks.
- $i$ represents the index of the $i$-th event in the incident.
- $j$ represents the index of the $j$-th character in the $i$-th event.
- $m_i$ is the maximum or total number of characters that can be processed in the $i$-th event. With :

MaxLength

$$= \max(\text{Length of } E_1, \text{Length of } E_2, \ldots, \text{Length of } E_n) \quad (7)$$

- $MaxLength$ represents the greatest number of characters found in any single event within the incident, ensuring that the maximum event length is selected as the standard size of a chunk.
- If a chunk does not reach the $MaxLength$ limit, it is padded and adjusted to this limit.

Finally, the total length of a cyber incident represents the sum of the lengths of all events or chunks within the incident:

$$T(C_{Incident}) = \sum_{i=1}^{t} T(E_i) = \sum_{i=1}^{t} T(C_i) \quad (8)$$

- $T_{(Incident)}$ represents the total length of the incident, calculated as the sum of the lengths of each event $E_i$ or chunk $C_i$.
- $t$ represents the total number of events or chunks in the incident. This total length determines how many tokens will be used to process the entire incident, ensuring that the chunking and tokenisation process adheres to the embedding model's constraint.

The following Figure 6 illustrates a visual representation of our context-specific chunking method applied to an incident knowledge base (Events):

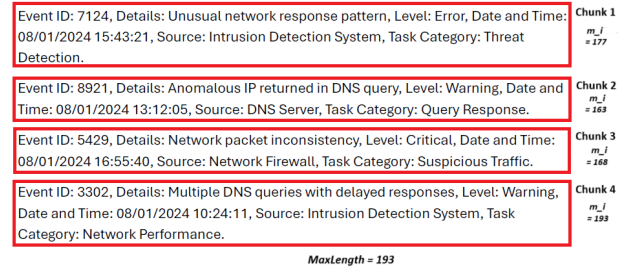

Fig. 6: DFIR Context-Specific Chunking

However, the practical application of this is shown in Listing 1, which provides a concise overview of the process and illustrates its implementation:

```python
# 'events' is a text containing multiple events
# Max length of a chunk ( Max number of characters
    within a chunk)
MaxLength = 208   # Set the fixed maximum length for
    each event/chunk (Maximum number of characters
    in a chunk)
chunks = []       # Store the chunks

# Split the events text by (.) to process each event
    separately
events = events.split(".")  # or ("/") - Split the
    document into events

# Loop over each split event
for event in events:
    # Remove extra spaces from each event for
        accurate characters calculation
    event = event.strip()

    # Check if the event is < than the max length,
        and pad if necessary
    if len(event) < MaxLength:
        # If the event is < than MaxLength, pad it
            with spaces
        event = event.ljust(MaxLength)

    # Add the event as a chunk
    chunks.append(event)  # Event is a chunk with
        the required length
```

Listing 1: Chunking code snippet

### C. Incident Events and DFIR Query Embedding

1) **Events**: Following the chunking process, this step focuses on embedding the chunks into high-dimensional vector representations. An embedding model is adopted to transform each chunk or event into a dense vector, which captures its semantic meaning and enables efficient similarity search and analysis. The first step involves passing an individual event $E_i$ through the embedding model, which first tokenises the event (i.e., splits it into tokens such as words or subwords) and then generates its vector $\mathbf{v}_{E_i}$.

$$\mathbf{v}_{E_i} = \text{Embed}(\text{Tokenise}(E_i)) \qquad (9)$$

$$\mathbf{v}_{E_i} = \text{Embed}(E_i) = \text{Embed}(\langle e_1, e_2, \dots, e_n \rangle) \qquad (10)$$

- The event $E_i$ consists of several attributes $e_1, e_2, \dots, e_n$, where each attribute $e_k$ represents a specific aspect or detail of the event (such as the date, description, etc.).

2) **DFIR Query**: The DFIR expert conducts automated timeline analysis by querying or instructing the framework. The query/instruction $Q_{DFIR}$ undergoes an embedding process similar to that of the event $E$. When the user inputs the query, it is first tokenised and then embedded, transforming it into a dense vector $\mathbf{v}_{Q_{DFIR}}$. The process is expressed as:

$$\mathbf{v}_{Q_{DFIR}} = \text{Embed}(\text{Tokenise}(Q_{DFIR})) \qquad (11)$$

### D. Context-Specific LLM-Powered RAG Agent for Timeline Analysis

1) **Events Retrieval**: The retrieval phase is initiated when the DFIR expert provides a query related to an incident. The query is embedded into a vector $v_{Q_{DFIR}}$, along with the vectors representing anomalous events $v_{E_i}$ stored in the external knowledge base $KB$. We represent the retrieval process of GenDFIR as:

$$\text{RAG}_{\text{DFIR}}(\mathbf{v}_{Q_{DFIR}}, \mathbf{v}_{E_i})$$
$$= \text{Retrieve}(\mathbf{v}_{Q_{DFIR}}, \{\mathbf{v}_{E_i}\}_{i \in R}, \text{cosine\_similarity}(\mathbf{v}_{Q_{DFIR}}, \mathbf{v}_{E_i}))$$

These embeddings allow the framework to semantically compare the input query with historical incident events. The similarity between the query vector $v_{Q_{DFIR}}$ and each event vector $v_{E_i}$ is computed using the cosine similarity function:

$$\text{cosine\_similarity}(\mathbf{v}_{Q_{DFIR}}, \mathbf{v}_{E_i}) = \frac{\mathbf{v}_{Q_{DFIR}} \cdot \mathbf{v}_{E_i}}{\|\mathbf{v}_Q\| \|\mathbf{v}_{E_i}\|} \qquad (12)$$

- $v_{Q_{DFIR}} \cdot v_{E_i}$ is the dot product of the query and event vectors, quantifying their alignment in the vector space.
- $\|v_{Q_{DFIR}}\|$ and $\|v_{E_i}\|$ are the Euclidean norms (magnitudes) of the respective vectors.

The cosine similarity identifies how closely the query aligns with the context of each anomalous event. This computation forms the basis for selecting the most relevant events. To ensure efficiency, only the top-$k$ events with the highest similarity scores are retrieved, as defined by:

$$\mathcal{R} = \text{TopK}\left(\left\{\frac{\mathbf{v}_{Q_{DFIR}} \cdot \mathbf{v}_{E_i}}{\|\mathbf{v}_{Q_{DFIR}}\| \|\mathbf{v}_{E_i}\|} \;\middle|\; i = 1, 2, \dots, t\right\}, k\right) \qquad (13)$$

- $\mathcal{R}$ represents the set of retrieved event vectors ranked by similarity scores. It plays a critical role in narrowing down the large pool of potential event to a focused subset of highly relevant anomalous events.
- $t$ denotes the total number of events in the knowledge base $KB$.
- $k$ is a hyperparameter that is defined and set manually based on the number of events within the knowledge base that are intended to be considered for analysis. The choice of $k$ directly impacts the quality and completeness of the DFIR timeline analysis, while an optimally defined $k$ equal to the exact number of events in the controlled knowledge base $KB$, ensures all evidence is included.

2) **DFIR Context-Specific Agent Workflow**: Once the initial set of relevant events has been retrieved, the DFIR RAG agent focuses on further filtering and refining these results based on contextual relevance. During this phase, the agent focuses on identifying and filtering the most contextually relevant events, ensuring that irrelevant evidence is excluded according to its assigned task as shown in Listing2.

```
DFIR_Agent_Prompt = """
You are a DFIR AI assistant, tasked
```

```
3  with analysing artefacts, correlating events,
       and producing a
4  coherent timeline of the incident. Base your
       answer on the
5  provided context and do not include additional
       information
6  outside of the context given.
7  """
8
9  DFIR_Agent = {
10     "role": "DFIR Timeline Analysis AI Assistant
       ",
11     "content": DFIR_Agent_Prompt,
12     "maxtokens": "To define"
13 }
```

Listing 2: Code snippet for GenDFIR Agent Prompt and Role

To augment the retrieval process, the framework adopts a matrix-based representation of event embeddings, defined as:

$$\mathcal{R} = \text{TopK}(\mathbf{V}_E \mathbf{v}_{Q_{DFIR}}^\top, k) \quad (14)$$

- $V_E$ is the matrix containing the embedding vectors of all events in the knowledge base $KB$.
- $v_{Q_{DFIR}}$ is the embedding vector of the query.
- $k$ is the predefined number of top events to retrieve, equal to the number of events in the controlled knowledge base.

This matrix-based approach allows for batch processing of embeddings, where similarity scores between the query and each event are computed simultaneously. To compute these scores, the query embedding $v_{Q_{DFIR}}$ is transposed (i.e., $v_{Q_{DFIR}}^\top$) so that it can be aligned with the event embeddings in the matrix $V_E$. This transposition ensures that the query vector is properly oriented for the dot product calculation with each event's embedding. The events in $\mathcal{R}$, being both relevant and contextually aligned with the DFIR query $Q_{DFIR}$, are then passed to the next stage of the framework, where they serve as the foundation for generating DFIR timelines and conducting further analysis by the LLM.

At this point, the agent not only retrieves the top-$k$ Evidence (relevant events) based on their similarity scores but also refines the results based on DFIR Timeline Analysis context-specific filtering.

3) **Timeline Analysis Generation**: In this phase, the framework utilises both the context provided by the retrieval process (Relevant Evidence $V_{RE_i}$) and the user's input $Q_{DFIR}$ to generate a timeline of anomalous events. The process involves multiple steps, including attention-based mechanisms, contextual enrichment by the LLM [56], and final timeline generation via a decoder-based LLM model. After retrieving the relevant evidence, the framework applies attention mechanisms. Attention scores are calculated between the query vector $v_{Q_{DFIR}}$ and the event vectors $v_{RE_i}$, allowing the model to weigh the relevance of each event. The attention score $\alpha_i$ is computed as

follows:

$$\alpha_i = \frac{\exp\left(\frac{\mathbf{v}_{Q_{DFIR}}^\top \mathbf{v}_{RE_j}}{\sqrt{d}}\right)}{\sum_{j=1}^k \exp\left(\frac{\mathbf{v}_{Q_{DFIR}}^\top \mathbf{v}_{RE_j}}{\sqrt{d}}\right)} \quad (15)$$

- $d$ is the dimension of the vectors (the number of features in each vector),
- $k$ is the number of the retrieved events.

This process amplifies the differences in similarity scores between events and the query, so events that are more similar to the query will have higher attention scores. This attention score is calculated using the dot product between $v_{Q_{DFIR}}$ and $v_{RE_i}$, followed by an exponential function. The exponential function serves two primary purposes in this context. First, it emphasises retrieved incident events $v_{RE_i}$ that are more relevant to the query $v_{Q_{DFIR}}$ by assigning them higher attention scores, which helps these events contribute more to the final timeline generation. Second, it ensures that the difference between similarity scores is amplified in a controlled manner, preventing events with slightly lower similarity from being overshadowed or overlooked. This balance allows the model to consider both highly relevant events and those with smaller, yet still meaningful, similarity and relevance to $v_{Q_{DFIR}}$.

Once the attention score $\alpha_i$ is calculated, the next step is to enrich the context of each event. The weighted sum of event vectors, adjusted by the attention scores, represents an initial context $c$ for the events, reflecting their weighted relevance to the query $v_{Q_{DFIR}}$:

$$\mathbf{c} = \sum_{i=1}^k \alpha_i \mathbf{v}_{RE_i} \quad (16)$$

The LLM further enriches this context by processing the weighted sum of events $c$ in light of its pre-existing training knowledge. For instance, Llama 3.1 has been extensively trained to recognise anomalous patterns, as well as publicly known cyber incident threats and anomalies. In this case, the context $c_{\text{final}}$ represents a nuanced, model-based interpretation of anomalous events, incorporating the LLM's understanding $c_{\text{extra}}$ of their significance based on patterns, relationships, and domain knowledge learned during training. This enriched context incorporates event correlation, where the LLM identifies and links related events based on temporal order, causality, or contextual associations.

For example, failed login attempts in a cyber incident might signal unauthorised access. The LLM correlates this event with subsequent related activities, such as privilege escalation or data exfiltration attempts, tying them to the same threat source. By analysing timestamps, anomaly severity, and underlying causes, the LLM constructs a logically sequenced timeline of events, capturing the progression and full scope of the incident. Without such automation, resolving incidents would require significant time and expertise, particularly for analysing complex event relationships and uncovering causal links. This process often demands considerable manual effort from specialists with deep technical knowledge. In contrast, GenDFIR leverages the LLM's extensive contextual understanding to accelerate and improve incident analysis, minimising reliance on manual intervention.

$$\mathbf{c}_{\text{final}} = Enrich\_Context(\mathbf{c}, \mathbf{c}_{\text{extra}}) \quad (17)$$

The final enriched context $c_{\text{final}}$, derived from the attention mechanism and LLM-based enrichment, is represented as a vectorised timeline $VTA$. This step involves integrating relevant evidence and user inputs to generate a numerical representation of the incident's timeline, as shown in Equation 17.

$$VTA = c_{\text{final}} \quad (18)$$

The final enriched context $c_{\text{final}}$, derived from the attention mechanism and LLM-based enrichment, is represented as a vectorised timeline $VTA$. This step involves integrating relevant evidence $v_{RE_i}$ and user inputs $v_{Q_{DFIR}}$ to generate the incident's timeline, as shown in Equation 18.

At this stage, $VTA$ has all relevant event information in a numerical form. It reflects the incident's sequence, correlation, and anomaly patterns across multiple dimensions (e.g., event features, timestamps, anomaly levels).

To make the timeline accessible and understandable to humans, $VTA$ must be decoded into a readable format. This is achieved by the decoder-based LLM model, which transforms the representation into a human-readable, logically structured timeline of events as in Equation 19:

$$\hat{TA} = \text{LLM\_Decoder}(VTA) \quad (19)$$

- Where $\hat{TA}$ denotes the model's generated or predicted timeline, in a human-readable format, suitable for further analysis, interpretations and decision-making by Investigators or DFIR Experts.

The GenDFIR workflow is depicted in Figure 7 and is summarised in the following algorithm 1:

## IV. GENDFIR IMPLEMENTATION

To properly implement the proposed GenDFIR framework, it is essential to establish clear distinctions regarding its intended purpose and scope. The GenDFIR framework is not designed to function as a conventional forensic tool that strictly adheres to legal evidentiary standards, nor is it intended for direct application in professional forensic cases that require full compliance with legal and regulatory frameworks at this stage of research. Instead, the framework's conceptualisation, development, and design serve as a foundational base, demonstrating the potential use of GenAI, specifically LLMs, within DFIR scenarios.

An important consideration for its correct implementation is the training knowledge of the LLM, especially the enriched context tied to the events in its knowledge base. LLMs raise significant concerns about the origin of their training

---

**Algorithm 1** GenDFIR

**Require:** DFIR Expert/Analyst Query or Instruction as input $Q_{DFIR}$ , Incident Knowledge Base $ID$

**Ensure:** Auto-Generated Timemline Analysis Report relevant to the incident $TA_{Report}$
1: Process Incident Document $ID$
2: Load LLM Model Llama-3.1 $llama - 3.1$
3: Load LLM Embedding Model 'mxbai-embed-large'
4: $Q_{DFIR}$ = "Conduct DFIR timeline analysis by examining the artefact, correlating events, and reconstructing the timeline of the cyber incident"
5: $EmbedModel$ = 'mxbai-embed-large'
6: $LLM = llama - 3.1$
7: $Chunk$ ($ID$)
8: $Tokenise$ ($E$, $Q_{DFIR}$)
9: $VE = Embed$ ($E$, $EmbedModel$)
10: $VQ_{DFIR} = Embed$ ($E$, $EmbedModel$)
11: $VRE = Agent_{DFIR}.Process(TopK, VQ_{DFIR}))$
12: $VTA = Enrich\_Context(VRE, VQ_{DFIR}, LLM)$
13: $TA_{Report} = LLM\_Decoder(VTA)$

---

data, which must come from legally compliant repositories in accordance with regulations such as the GDPR in Europe. For instance, Meta has chosen not to release its latest models like Llama-3.2 [57] in the EU due to regulatory uncertainties, particularly regarding the data used for its training as stated by *"The Guardian"* [58]. In this research, Llama 3.1 is implemented in a zero-shot setting, which aligns with UK GDPR and security standards. No fine-tuning has been applied, ensuring the framework relies entirely on the model's original capabilities.. Furthermore, the data used in the experiment is entirely synthetic, generated to simulate realistic scenarios while maintaining control over results monitoring and analysis. However, further considerations and concerns are addressed in the following lines, along with the components of the framework, the experimental environment, and the case studies used for testing:

1) **Variability in DFIR Timeline Analysis Practices and GenDFIR:** It is essential to recognise the diversity of methodologies and procedures within the DFIR industry. Each organisation or entity may adopt different investigative workflows and reporting formats, depending on the specific nature of the investigation. Broadly, DFIR processes can be divided into two primary domains: DF, which focuses on the preservation, collection, and analysis of digital evidence, and IR, which centres on the containment, eradication, and recovery from cyber incidents.

Organisations vary in their approach to timeline analysis during investigations, as enterprise-focused IR differs from legal forensics in terms of the depth of analysis, objectives, and reporting. In enterprise environments, timeline analysis often prioritises speed and operational impact, emphasising the rapid identification of critical events and resolution steps, sometimes with less emphasis on maintaining a formal chain of custody. In contrast, legal forensic investigations place greater importance on the preservation and reconstruction of timelines, ensuring that all digital events are accurately sequenced and documented to meet evidentiary standards for potential legal proceedings.

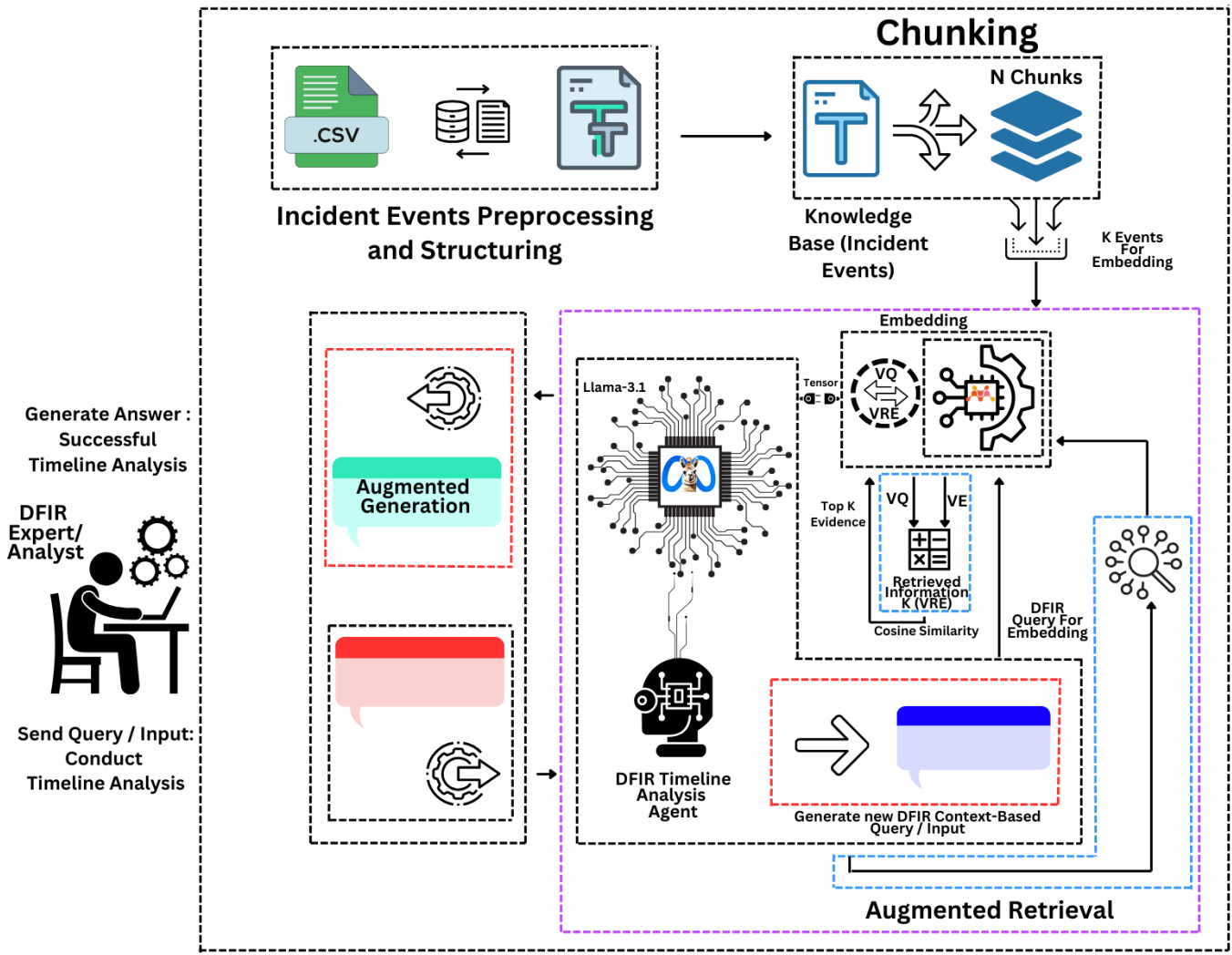

Fig. 7: GenDFIR Framework Workflow

The reporting mechanisms for timeline analysis also differ. IR reports typically concentrate on delivering actionable insights—such as the timing of compromise, propagation, and remediation steps—often to guide immediate decision-making. These reports prioritise clarity and immediacy over legal formalities. In contrast, forensic timeline analysis reports must adhere to strict documentation standards, ensuring that every event is precisely tracked and presented in a manner that preserves the integrity of the investigation for legal scrutiny [2]. This involves rigorous validation of timestamps, metadata, and cross-referencing multiple data sources to reconstruct a legally defensible timeline of events.

The GenDFIR framework is not a one-size-fits-all solution, as DFIR practices differ across organisations. The framework is designed as a flexible approach to integrate GenAI models into DFIR workflows, supporting experts in improving the efficiency of investigations. The framework focuses on tasks such as log analysis, anomaly detection, and rapid evidence identification, with the purpose of providing a context-rich timeline analysis to help DFIR experts make informed decisions more quickly.

2) **Large Language Models in GenDFIR:** The practical application of LLMs within DFIR tasks is still novel. Recent advancements in the use of LLMs across various fields have generated interest, but they are far from being fully reliable tools for DFIR investigations [59]. Several challenges must be acknowledged, including issues of hallucination, precision, and limitations related to input/output token lengths, which can restrict the model's ability to process large datasets typical in DFIR cases.

LLMs are powerful tools for automating certain aspects of data analysis, summarisation, and even generating incident reports. However, they are not yet capable of performing highly specialised forensic tasks without human oversight.

3) **Data in GenDFIR:** The LLM used in GenDFIR operates in a zero-shot mode, relying solely on its pre-existing training data to generate general-purpose DFIR Timeline Analysis reports. While this approach works for broad applications, it is insufficient for the specific and dynamic needs of DFIR, which requires real-time recognition and processing of incident-specific data. This is particularly important during the phase of enriching the context of incident events, where the effectiveness of enrichment and interpretation depends on the LLM's pre-existing knowledge and its ability to process relevant data. Furthermore, each DFIR case involves unique and sensitive information, often shaped by regulatory frameworks, organisational policies, and legal requirements. As a result, incident data is highly diverse, reflecting variations in breach detection and response mechanisms.

This diversity highlights a significant gap between generalised knowledge and the nuanced demands of DFIR. Bridging this gap requires moving beyond generic outputs, such as the Timeline Analysis report generated by GenDFIR, and integrating incident-specific data into the model. This can be achieved by training LLMs in secure, controlled environments using curated datasets derived from real-world incidents. Such an approach enables LLMs to adapt to the distinct characteristics of various DFIR scenarios, improving their ability to identify patterns and respond effectively across different contexts.

However, the integration of real-world data into LLM training raises important privacy, legal, and ethical considerations. DFIR investigations often involve sensitive information, including personal data and proprietary organisational content. The use of incident-specific data must adhere to data protection laws, organisational consent policies, and robust security measures to safeguard against breaches during both model training and inference. These safeguards are critical to ensuring confidentiality while maximising the value of such data for future applications.

Moreover, current LLMs face limitations in adapting to new or evolving threats outside their original training scope. Without regular updates and retraining on fresh, ethically and legally sourced datasets, their performance in DFIR scenarios may degrade over time.

### A. GenDFIR Components

The following are the components integrated into GenDFIR:

- **LLM Model**: Llama-3.1 a powerful model developed by Meta, evaluated on a range of standard benchmarks and human evaluations. The rationale behind selecting this model is its open-source nature and training on cybersecurity-related content [43]. This offers significant advantages for GenDFIR, as it suggests that the LLM can analyse and potentially recognise anomalous events, providing insights into cyber incidents. The model is available in different versions with varying parameter sizes, including 405B, 70B, and 8B parameters. At an early stage, we adopt the 8B model to reduce hallucinations and for system prototyping.

- **Embedding Model**: mxbai-embed-large, a Mixedbread release, with a maximum token limit of 512 and a maximum dimension of 1024. The model performed highly on the Massive Text Embedding Benchmark (MTEB) [60]. The reason for opting for this model is its capabilities in retrieval and semantic textual similarities. Additionally, it is effective for scoring methods during the retrieval process, such as cosine similarity, adopted in GenDFIR.

- **DFIR RAG Agent**: Autonomous Context-Augmented Agent, powered by Llama-3.1. Employed in GenDFIR for context-based retrieval and output generation. Operating autonomously as an expert in Cybersecurity and DFIR, specialising in DFIR Timeline Analysis and tasked with identifying events and information related to cyber incidents from our knowledge base containing cyber incident events.

### B. Experiment Environment

GenDFIR and the LLM that empowers it are designed to operate locally. The data used for the experiment are synthetic and represent a variety of cyber incidents that mimic the characteristics of Windows event logs, phishing scenarios (emails), and other cyber incidents, all presented in their textual format and in English. The following outlines the details of the experimental environment and conditions in which GenDFIR was tested:

1) **Python Libraries:** Our code was written in Python. The primary libraries used are `PyTorch` for tensor operations, `NumPy` for numerical computations such as cosine similarity, and `pandas` for data manipulation and handling embeddings.

2) **LLM Response Temperature:** The temperature of the core LLM (`Llama3.1:8b`) is set to a very low value (`temperature = 0.1`) to reduce perplexity and control prediction variability. This limits the generation of unnecessary content, enabling the framework to focus more effectively on the knowledge retrieved from the knowledge base, rather than relying on the LLM's additional and irrelevant internal knowledge.

3) **Max Output Tokens:** is set to (`maxtokens=2000`) to work conveniently with the embedding model (`mxbai-embed-large:335m`) and the LLM (`llama3.1:8b`) specifications.

4) **Number of Completions:** is set to a low value (n=1) to make the LLM's generation more deterministic and to select the optimal word predictions.

```
1  answer = client.chat.completions.create(
2      model=LlmModel,
3      messages=[{"role": "system", "content":
          prompt}],
4      maxtokens=2000,
5      temperature=0.1,
6      n=1,
7  )
```

Listing 3: GenDFIR code snippet Visualisation

5) **Parallel Computing:** For optimised tensor computation, CUDA is used to run the tensors on the GPU.

```
1  GPU Model: NVIDIA GeForce RTX 4060
2  CUDA Version: 12.6
3  Total VRAM: 8GB
```
Listing 4: CUDA and GPU

### C. Datasets Elaboration

The experiments in this study are based on synthetic scenarios designed to mimic real-world cyber incidents while preserving privacy, consent, and confidentiality. Each scenario has been customised to meet the specific requirements of the experimental setup (`maxtokens=2000`) and (`llama-3.1:8b`) running locally, ensuring compatibility with its design and objectives. The following Table IV outlines the descriptions and details of the incident scenarios used in the experiments:

## V. RESULTS, EVALUATION AND DISCUSSION

### A. Results

The results of our research are presented and found in https://github.com/GenDFIR/GenDFIR [62]. These results consist of the Timeline Analysis Reports generated by GenD-FIR. As the reports are quite large in size, this paper showcases only one example, with others available through the previously mentioned link.

To generate the reports, it is crucial to first visualise the knowledge base. For this paper, we have chosen to showcase the 'Unauthorised Access' scenario. Table V illustrates our knowledge base, which also serves as the ground truth. This knowledge base is uploaded to the framework to function as its dynamic memory, enabling the retrieval of events as needed (either simultaneously or at different times) depending on the context:

The prompt/query/instruction provided by the DFIR expert is shown in Table VI:

The DFIR RAG Agent task and prompt is in Table VII:

The output is given in VIII:

As shown, the report contains an analysis of the artefacts, which are the events in the knowledge base, their correlations, and their interpretation in relation to the incident. Additionally, context enrichment has been applied to the incident events to enhance their interpretability. The framework also provides additional knowledge, such as identifying anomalous events and trends, root causes, and mitigation solutions. This reflects the role of the agent acting as a DFIR assistant, which, in real-life scenarios, provides such information during a cyber incident.

However, these results cannot be fully deemed relevant without further evaluation, which is presented in the following section.

### B. Evaluation

Since GenDFIR is a context-specific framework that employs LLMs in a zero-shot setting, we propose a tailored evaluation approach focused on assessing the framework's outputs, including the effectiveness of retrieval, evidence identification, and extraction. Traditionally, LLMs are evaluated using established benchmarks such MMLU, which measure general-purpose language understanding and reasoning. For example, Llama-3.1 has been benchmarked by Meta, demonstrating its reliability not only for general-purpose applications but also for general cybersecurity contexts [43]. However, such benchmarks do not adequately reflect the requirements of highly specialised domains like DFIR.

Metrics like F1-score, precision, and recall, originally designed for numerical or classification tasks, fail to capture the complexity of evidence retrieval and context-specific analysis in this context. To address this, we propose and adapt metrics such as accuracy, relevance, exact match, and top-k evidence retrieval to suit the unique demands of this framework. The design and application of these metrics are described in the following sections:

1) **Accuracy:** Accuracy in GenDFIR is determined by evaluating the generated timeline analysis reports to identify and verify the factual content, as well as to assess how and to what extent these reports can be useful and reliable in assisting DFIR experts. This process involves cross-referencing the generated information with a verified knowledge base (Incident Events) to ensure correctness and relevance while confirming that interpretations are logical and reasonable. The reports are divided into two sections:

   a) **Knowledge Base (Events) Facts Section**: This section includes artefact analysis, timeline analysis, event correlation, and timeline reconstruction. These components are derived from facts retrieved from the knowledge base and enriched with contextual information by the LLM (Llama-3.1).

   b) **Additional Insights Facts Section:** This section provides supplementary information, including mitigation strategies, recommendations, and other relevant insights. These outputs are generated using the general knowledge of the LLM (Llama-3.1) and its contextual application to the knowledge base (Incident Events).

The accuracy of GenDFIR reports is evaluated using a proposed equation designed to quantify the proportion of verified correct facts relative to the total number of facts generated. The equation is defined as:

$$\text{Accuracy} = \frac{\text{Overall Correct Facts}}{\text{Overall ( Correct Facts + Incorrect Facts)}} \quad (20)$$

Where according to Table IX:

- **Overall Corerct Facts:** Are the total number of accurate facts obtained by combining the correct facts from the incident knowledge base with the correct facts generated by the LLM, as shown in Table IX
- **Correct Facts:** Are facts that are verified and deemed accurate individually from both the incident knowledge base and the LLM-generated facts
- **Incorrect Facts:** Similarly, facts that are inaccurate and identified separately from each source.

TABLE IV: Cyber Incident Scenarios

| Scenario | Description | Number of Events (Chunks) | Max Chunk Length (Characters - MaxLength) per Event | Event Splitter |
|---|---|---|---|---|
| SYN Flood | This is a SYN flood attack, where unusual network events disrupted standard operations. The anomalies were characterised by a high volume of Synchronise (SYN) requests, causing intermittent service degradation across the network. Data was collected from firewalls, network scanners, and Intrusion Detection Systems (IDS). The analysis focus on critical attributes such as event ID, details, level, timestamps, source, task category, and affected devices to assess the nature of the attack and identify potential threats or operational issues.<br><br>• Each event log represents a chunk. | 30 | 210 | " . " |
| Rhino Hunt (Inspired from [61]) | This scenario is inspired by the well-known "Rhino Hunt" incident, but in this case, it involves the illegal exfiltration of copyrighted rhino images. An unauthorised individual accessed the company's FTP server and stole twelve protected images. The investigation traced the exfiltration to a device within the company's internal network, with the stolen data directed to an external IP address. Forensic analysis revealed that the user associated with the IP address possessed additional images matching the stolen ones. The collected data included images that met specific metadata criteria, including camera model, artist, and copyright details.<br>- The images used in this scenario are AI-generated (using DALL-E 3) for the purpose of ensuring copyright compliance and consent. The metadata of the images has been modified to align with the scenario description and can be extracted and viewed using the Metaminer module found on [62].<br><br>• The events in this scenario represent log entries where the context of the image has been added at the beginning of each event to enrich the entry with additional context. | 8 | 500 | " / " |
| Phishing Email - 1 | This scenario represents a phishing attack where an employee was targeted by emails impersonating a security service. The organisation's policy prohibits communication with untrusted domains, permitting only interactions with verified sources. All suspicious emails were collected for analysis, focusing on domains, sender and receiver details, IP addresses, email content, and timestamps to determine the nature of the attack.<br><br>• Each single Email represents a chunk. | 15 | 725 | " / " |
| Phishing Email - 2 | Same as the previous scenario, this is a phishing attack where an employee was targeted by emails impersonating a trusted support service, attempting to deceive the employee into verifying account information. The organisation's policy restricts communication with unverified domains and permits only trusted sources. All suspicious emails received during the suspected phishing period were collected for analysis, focusing on domains, sender and receiver details, IP addresses, email content, and timestamps to assess the nature of the attack.<br><br>• Each single Email represents a chunk. | 20 | 500 | " / " |
| DNS Spoof | This incident involves a DNS spoofing scenario where multiple event logs were collected from various devices, including Windows event logs, DNS server logs, firewall records, and network traffic monitoring tools. Irregularities such as delayed DNS responses, inconsistent resolutions, and unexpected outbound traffic were identified, triggering alerts from the Intrusion Detection System (IDS) and performance monitoring tools.<br><br>• Each event log represents a chunk. | 23 | 200 | " . " |
| Unauthorised Access | This scenario is an unauthorised access attempt detected by an intrusion detection system (IDS). The system flagged multiple access attempts from a blacklisted IP address, which was not authorised for any legitimate activity within the network. The collected data, including warnings, errors, and critical alerts, provided the basis for further investigation into the potential breach.<br><br>• Each event log represents a chunk. | 25 | 208 | " . " |

Based on extensive monitoring and assessment of the Timeline Analysis reports and results, as illustrated in Table IX, minor incident events found in the generated report, along with their interpretation and retrieval, have been identified as incorrect in some scenarios. For instance, in Phishing Email-1, it was reported that *"2. Follow-up emails attempting to get Michael to verify his account through a link (10:45 AM, 03:00 PM, and 04:30 PM)"* is incorrect, as the time (04:30 PM) does not exist and is not mentioned in the incident events knowledge base. On the other hand, the additional knowledge generated by the LLM was all correct. For example, in the SYN-FLOOD scenario, the key discoveries generated and found in the Timeline report were relevant **Key Discoveries** * Multiple SYN flood attacks detected throughout the day. * High volume of SYN packets and excessive packet volumes exceeded thresholds. * Regular network performance checks revealed potential issues and were successfully cleared. * Even though these discoveries and knowledge are general knowledge related to any type of SYN flood attack, they were accurate in this context.

The following graph 8 illustrates the accuracy rate for each scenario, calculated according to the previously given equation and results found in Table IX:

By calculating the average accuracy across all scenarios, the overall accuracy of GenDFIR is 95.52%.

| Knowledge Base (Incident Events - Ground Truth Data) | Event ID: 4625, Details: Logon attempt failed, Level: Warning, Date and Time: 2016/09/15 02:15:05, Source: Windows Security, Task Category: Logon. |
| --- | --- |
| | Event ID: 5038, Details: Integrity of system file failed, Level: Error, Date and Time: 2016/09/15 02:16:37, Source: System, Task Category: System Integrity. |
| | Event ID: 4723, Details: Attempt to change account password, Level: Warning, Date and Time: 2016/09/15 02:17:05, Source: Windows Security, Task Category: Account Management. |
| | Event ID: 4625, Details: Logon attempt failed, Level: Warning, Date and Time: 2016/09/15 02:17:21, Source: Windows Security, Task Category: Logon. |
| | Event ID: 1102, Details: Audit log cleared, Level: Critical, Date and Time: 2016/09/15 02:18:05, Source: Security, Task Category: Audit Logs. |
| | Event ID: 4625, Details: Logon attempt failed, Level: Warning, Date and Time: 2016/09/15 02:18:55, Source: Windows Security, Task Category: Logon. |
| | Event ID: 4769, Details: Kerberos service ticket request failed, Level: Warning, Date and Time: 2016/09/15 02:20:12, Source: Security, Task Category: Credential Validation. |
| | Event ID: 4625, Details: Logon attempt failed, Level: Warning, Date and Time: 2016/09/15 02:21:00, Source: Windows Security, Task Category: Logon. |
| | Event ID: 4771, Details: Pre-authentication failed, Level: Warning, Date and Time: 2016/09/15 02:22:30, Source: Security, Task Category: Credential Validation. |
| | Event ID: 4648, Details: Logon attempt with explicit credentials, Level: Warning, Date and Time: 2016/09/15 02:23:35, Source: Windows Security, Task Category: Logon. |
| | Event ID: 4724, Details: Password reset attempt, Level: Warning, Date and Time: 2016/09/15 02:24:12, Source: Windows Security, Task Category: Account Management. |
| | Event ID: 5031, Details: Firewall service encountered an error, Level: Warning, Date and Time: 2016/09/15 02:24:45, Source: Security, Task Category: Firewall. |
| | Event ID: 4776, Details: Logon failure due to incorrect credentials, Level: Warning, Date and Time: 2016/09/15 02:27:15, Source: Windows Security, Task Category: Credential Validation. |
| | Event ID: 4771, Details: Pre-authentication failed, Level: Warning, Date and Time: 2016/09/15 02:30:01, Source: Security, Task Category: Credential Validation. |
| | Event ID: 4625, Details: Logon attempt failed, Level: Warning, Date and Time: 2016/09/15 02:30:45, Source: Windows Security, Task Category: Logon. |
| | Event ID: 4723, Details: Attempt to change account password, Level: Warning, Date and Time: 2016/09/15 02:31:55, Source: Windows Security, Task Category: Account Management. |
| | Event ID: 5038, Details: Integrity of system file failed, Level: Error, Date and Time: 2016/09/15 02:32:07, Source: System, Task Category: System Integrity. |
| | Event ID: 4625, Details: Logon attempt failed, Level: Warning, Date and Time: 2016/09/15 02:33:15, Source: Windows Security, Task Category: Logon. |
| | Event ID: 4648, Details: Logon attempt with explicit credentials, Level: Warning, Date and Time: 2016/09/15 02:34:15, Source: Windows Security, Task Category: Logon. |
| | Event ID: 4771, Details: Pre-authentication failed, Level: Warning, Date and Time: 2016/09/15 02:34:30, Source: Security, Task Category: Credential Validation. |
| | Event ID: 1102, Details: Audit log cleared, Level: Critical, Date and Time: 2016/09/15 02:35:00, Source: Security, Task Category: Audit Logs. |
| | Event ID: 4625, Details: Logon attempt failed, Level: Warning, Date and Time: 2016/09/15 02:35:12, Source: Windows Security, Task Category: Logon. |
| | Event ID: 4776, Details: Logon failure due to incorrect credentials, Level: Warning, Date and Time: 2016/09/15 02:36:15, Source: Windows Security, Task Category: Credential Validation. |
| | Event ID: 5031, Details: Firewall service encountered an error, Level: Warning, Date and Time: 2016/09/15 02:36:30, Source: Security, Task Category: Firewall. |
| | Event ID: 4625, Details: Logon attempt failed, Level: Warning, Date and Time: 2016/09/15 02:37:45, Source: Windows Security, Task Category: Logon. |

TABLE V: Incident Knowledge Base

| DFIR Expert - Analyst Query | Conduct DFIR timeline analysis by examining the artefact, correlating events, and reconstructing the timeline of the cyber incident |
| --- | --- |

TABLE VI: DFIR Expert Query/Instruction

2) *Relevance:* In this context, we define the relevance metric as the extent to which the answers provided by the framework are pertinent. To measure this, we have crafted 20 DFIR context-specific prompts (available in [62]) in the form of questions for each scenario related to the incident. These prompts are characterised by their focus on various aspects, including sentiment (e.g., *"What is the overall sentiment of the image described in Event 2 (Image b)?"* - Rhino Hunt Scenario), intention (e.g., *"What specific action is requested by GlobalBank in the initial email?"* - Email Phishing-1 Scenario), deep analysis (e.g., *"Analyze the severity levels associated* with the events involving SYN flood attacks. List the events and their corresponding severity levels."* - SYN-FLOOD Scenario), as well as retrieval, prediction, and insights. According to the generated answers, GenDFIR failed to provide correct responses for certain questions (DFIR prompts). For instance, in the SYN Flood scenario, GenDFIR failed on Prompt 4; in the Rhino Hunt scenario, Prompt 13 was answered partially correctly (50%); and in the Phishing Email 1 scenario, GenDFIR failed on both Prompts 11 and 12. The correctness and incorrectness of these answers were compared and cross-referenced with the incident events knowledge base, which is considered

| | |
|---|---|
| **GenDFIR RAG Agent** | **Role:** DFIR Timeline Analysis AI Assistant |
| | **Agent Prompt:** You are a DFIR AI assistant, tasked with analysing artefacts, correlating events, and producing a coherent timeline of the incident. Base your answer on the provided context and do not include additional information outside of the context given. |

TABLE VII: DFIR Agent Prompt and Role

| | |
|---|---|
| **GenDFIR output (Powered by Llama-3.1)** | **\*\*Incident Analysis\*\*** |
| | I have analyzed the given event IDs to identify anomalous events, conducted a timeline analysis, correlated between them, reconstructed the events timeline, and provided precise mitigation solutions. |
| | **\*\*Event Timeline Reconstructed\*\*** |
| | 1. **\*\*2016/09/15 02:16:37\*\***: Event ID 5038 - Integrity of system file failed (Source: System) |
| | 2. **\*\*2016/09/15 02:17:05\*\***: Event ID 4723 - Attempt to change account password (Source: Windows Security) |
| | 3. **\*\*2016/09/15 02:17:21\*\***: Event ID 4625 - Logon attempt failed (Source: Windows Security) |
| | 4. **\*\*2016/09/15 02:20:12\*\***: Event ID 4769 - Kerberos service ticket request failed (Source: Security) |
| | 5. **\*\*2016/09/15 02:22:30\*\***: Event ID 4771 - Pre-authentication failed (Source: Security) |
| | 6. **\*\*2016/09/15 02:24:12\*\***: Event ID 4724 - Password reset attempt (Source: Windows Security) |
| | 7. **\*\*2016/09/15 02:24:45\*\***: Event ID 5031 - Firewall service encountered an error (Source: Security) |
| | 8. **\*\*2016/09/15 02:27:15\*\***: Event ID 4776 - Logon failure due to incorrect credentials (Source: Windows Security) |
| | 9. **\*\*2016/09/15 02:30:01\*\***: Event ID 4771 - Pre-authentication failed (Source: Security) |
| | 10. **\*\*2016/09/15 02:31:55\*\***: Event ID 4723 - Attempt to change account password (Source: Windows Security) |
| | 11. **\*\*2016/09/15 02:32:07\*\***: Event ID 5038 - Integrity of system file failed [duplicated] (Source: System) |
| | 12. **\*\*2016/09/15 02:34:30\*\***: Event ID 4771 - Pre-authentication failed [duplicated] (Source: Security) |
| | 13. **\*\*2016/09/15 02:36:15\*\***: Event ID 4776 - Logon failure due to incorrect credentials [duplicated] (Source: Windows Security) |
| | 14. **\*\*2016/09/15 02:36:30\*\***: Event ID 5031 - Firewall service encountered an error [duplicated] (Source: Security) |
| | 15. **\*\*2016/09/15 02:37:45\*\***: Event ID 4625 - Logon attempt failed [duplicated] (Source: Windows Security) |
| | **\*\*Anomalous Events and Trends\*\*** |
| | 1. Multiple attempts to change account password (Event IDs 4723) between 2:17 AM and 2:31 AM. |
| | 2. Repeated pre-authentication failures (Event ID 4771) between 2:22 AM and 2:34 AM. |
| | 3. Logon attempt failures (Event IDs 4625, 4627) multiple times between 2:17 AM and 2:37 PM. |
| | 4. Firewall service encountered errors (Event ID 5031) at different intervals between 2:24 AM and 2:36 PM. |
| | **\*\*Root Cause Analysis\*\*** |
| | These events suggest a possible malicious action or an automated script that was attempted to gain unauthorized access. The repeated failures may indicate a faulty password reset attempt or other security measures that could not be bypassed. |
| | **\*\*Mitigation Solutions\*\*** |
| | 1. **\*\*Enhance Security Measures\*\***: Immediate updates and patches should be applied to counter vulnerabilities. |
| | 2. **\*\*Alert System Deployment\*\***: Activate systems capable of detecting such attempts in real-time to prevent further occurrences. |
| | 3. **\*\*Audit Logging Review\*\***: Review all logging mechanisms to determine if more detailed records are needed. |
| | 4. **\*\*Training Session\*\***: Update security training to emphasize common attack vectors and prevention strategies. |
| | **\*\*Recommendations\*\*** |
| | 1. The security team needs to review the system's configuration for any vulnerabilities that could have been exploited. |
| | 2. Investigate these events further with a focus on identifying potential weaknesses the attacker might have used. |
| | 3. Conduct in-depth audits of systems, configurations, and account management policies to prevent similar incidents from recurring. While I've reconstructed the timeline and identified possible vulnerabilities, actual mitigation efforts should be handled by security teams who can take into consideration all necessary factors within your organization. |

TABLE VIII: GenDFIR Output

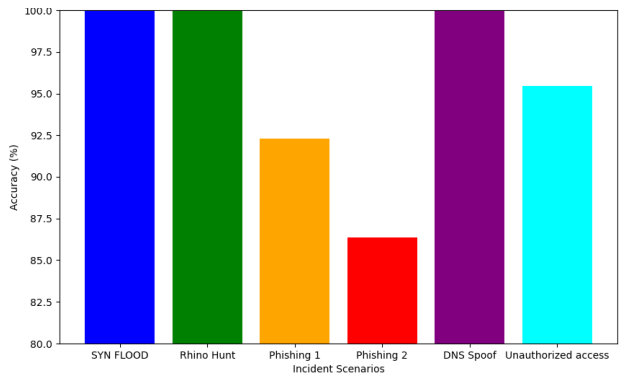

Fig. 8: Overall Accuracy of GenDFIR Timeline Analysis reports for different Cyber Incident Scenarios.

TABLE IX: Timeline Analysis Report Facts

| Scenarios | Overall Timeline Analysis Report Facts | Incident Knowledge Base | LLM Facts (All correct) |
|---|---|---|---|
| **SYN Flood** | 20 | - Correct: 17<br>- Incorrect: 0 | 03 |
| **Rhino Hunt** | 16 | - Correct: 08<br>- Incorrect: 0 | 08 |
| **Phishing Email - 1** | 13 | - Correct: 09<br>- Incorrect: 1 | 03 |
| **Phishing Email - 2** | 22 | - Correct: 13<br>- Incorrect: 03 | 06 |
| **DNS Spoof** | 19 | - Correct: 14<br>- Incorrect: 0 | 05 |
| **Unauthorized Access** | 22 | - Correct: 14<br>- Incorrect: 01 | 07 |

the ground truth data. The detailed failures and successes are shown in the graph in Figure 9:

The overall relevance rate of GenDFIR in this case is 94.51%.

3) **Exact Match:** The EM metric in GenDFIR is designed to evaluate the framework's ability to retrieve precise and granular information from the knowledge base. It tests the framework's performance by posing DFIR questions that require exact matches to the ground truth data. This includes specific details such as timestamps (e.g., *"What was the exact time when Michael Davis responded to the first email from the GlobalBank Security Team?"* and descriptions (e.g., *"In Event 5, what was the description of the alligator?)"*, along with other relevant queries. Results found in [62] show that GenDFIR has performed well, aligning perfectly with the ground truth. GenDFIR successfully passed all checks without any errors. Therefore, the EM in this case is rated at 100%, as all 20 prompts were answered correctly.

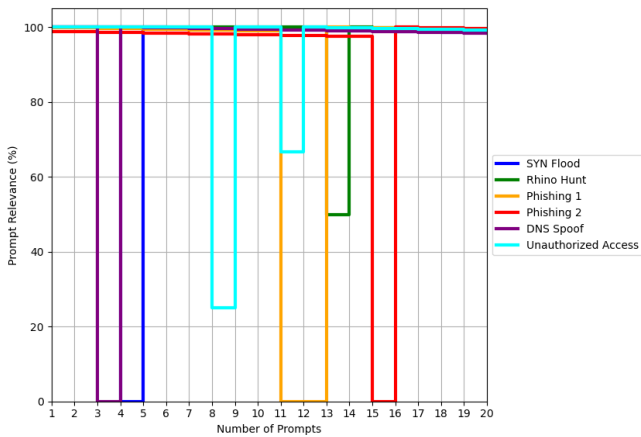

Fig. 9: Prompts Relevance of GenDFIR Timeline Analysis reports for different Cyber Incident Scenarios.

4) *Evidence (Top-k)*: DFIR Timeline Analysis is primarily focused on identifying evidence through the analysis of digital artefacts, with the correlation between events linked to the relevance of evidence deemed pertinent to the incident. To facilitate real-time visualisation of evidence identification during inference, we developed a script, illustrated in Listing 4, which uses the UMAP library.

- **UMAP Library Adaptation for GenDFIR:** The UMAP library contains algorithms used for dimensionality reduction and visualisation, enabling the identification of the closest data points in a 2D vector space. These algorithms work by calculating the distances between data points, selecting the nearest neighbours, and generating a graphical representation that illustrates how data points are related, preserving both local and global structures [63]. In the context of GenDFIR, we developed a script using this library to visualise the relationship between features from the vectorised user input $VQ_{DFIR}$ and their corresponding features in the embedded knowledge base (Incident Events $VE$). This process helps identify how input data from the DFIR expert aligns with relevant incident events.

  It is important to note that evidence identification in this scenario depends on the input provided by the user, who determines which events or attributes should be considered as evidence. In other words, the evidence is solely based on the query prompts issued by the DFIR expert, whose suspicious assessment guides the process. As discussed in the methodology, GenDFIR integrates algorithms based on cosine similarity and a DFIR-specific agent to assess relevance. In UMAP, we added a heat parameter to visualise the similarity during inference between the query from the DFIR expert and the events embeddings from the knowledge base. This heat parameter ranges from 0 to 1, where a value closer to 1 indicates stronger similarity between the query and the events embeddings, and a value closer to 0 indicates weaker similarity. The colour intensity in the visualisation reflects this similarity, with higher values (closer to 1) represented by warmer colours, signifying that the events are more relevant to the query. In addition to this, the script leverages the traditional functionality of the UMAP algorithm, which clusters data points according to how similar they are to each other.

To interpret the graphs generated by UMAP, data points that are close to each other indicate higher contextual similarity. During inference, events deemed relevant to the query will be positioned near one another in the visualisation. The heat parameter, represented by colour variation, reflects the degree of similarity between the data points, with warmer colours indicating a stronger match to the expert's query.

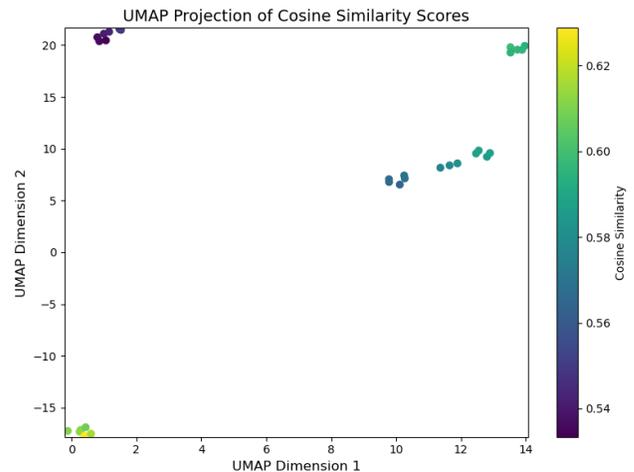

Fig. 10: SYN Flood - Evidence (Top K=5)

In the code, the `calculate_cosine_similarity` function computes the cosine similarity between the vectorised input $VQ_{DFIR}$ and the knowledge base embeddings $VE$. The `visualise_cosine_similarity` function then applies UMAP to reduce these similarity scores to a 2D space, where proximity indicates stronger similarity between events. The resulting visual representation enables the DFIR expert to better understand the alignment between the query and the knowledge base.

```
1 def calculate_cosine_similarity(VQ, VE):
2     """
3     Calculate cosine similarity between the input
        and Knowledge Base embeddings.
4     Args:
5     - VQDFIR (list or np.array): Embedding of the
        query.
```

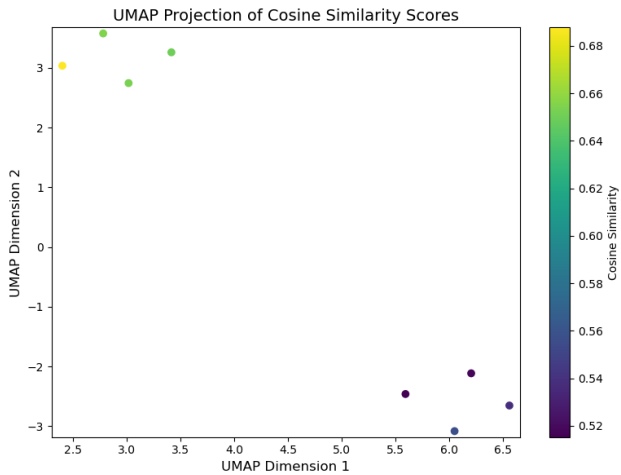

Fig. 11: Rhino Hunt - Evidence (Top K=4)

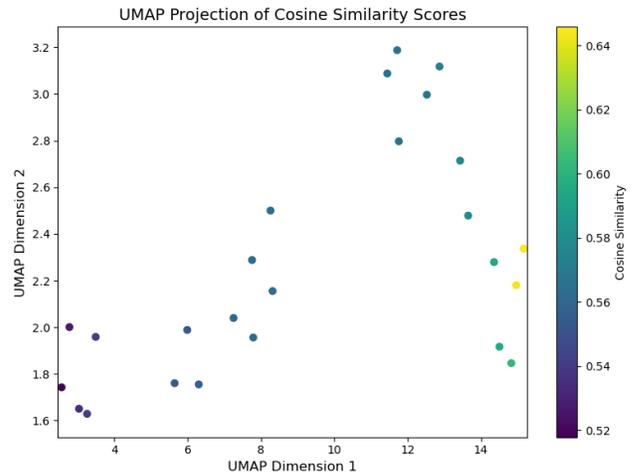

Fig. 14: Unauthorised Access - Evidence (Top K=25)

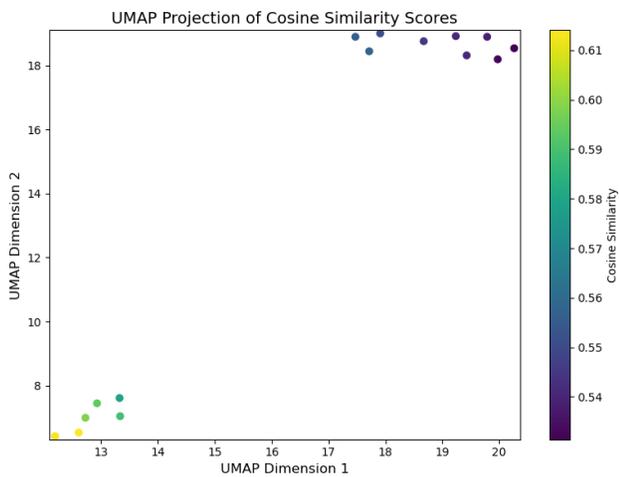

Fig. 12: Phishing Email 1 - Evidence (Top K=6)

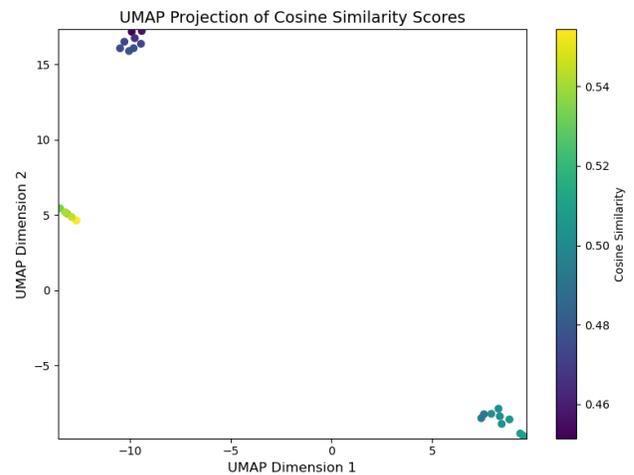

Fig. 15: DNS Spoof - Evidence (Top K=6)

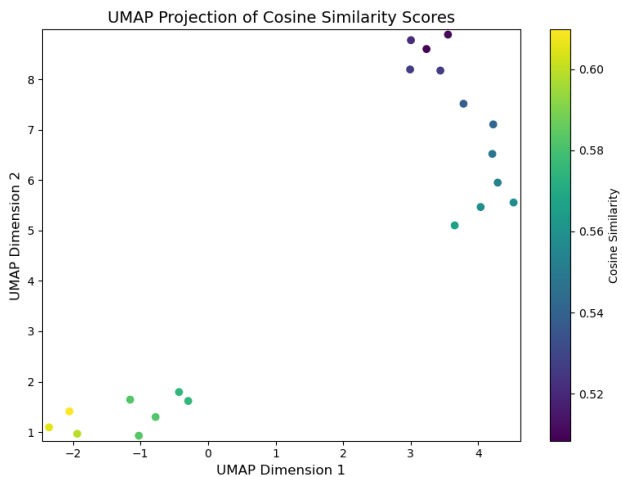

Fig. 13: Phishing Email 2 - Evidence (Top K=7)

```
10    cos_scores = torch.cosine_similarity(torch.
          tensor(VQ).unsqueeze(0), VE
11    return cos_scores
12
13 def visualise_cosine_similarity(cos_scores):
14    # Convert cosine similarity scores to 2D
15    scores_2d = np.array(cos_scores).reshape(-1, 1)
16    # Apply UMAP for dimensionality reduction
17    embedding = umap.UMAP(n_neighbors=5, min_dist
          =0.1, random_state=42).fit_transform(scores_2d)
18    # Create a scatter plot using Matplotlib
19    scatter = plt.scatter(embedding[:, 0], embedding
          [:, 1], c=scores_2d.flatten(), cmap='viridis')
20    plt.show()
21
22 cos_scores = calculate_cosine_similarity(VQDFIR, VE)
23 # Visualise the cosine similarity scores
24 visualise_cosine_similarity(cos_scores)
```

Listing 5: Code snippet for Calculating and Visualising Similarity

Figures 10, 11, 12, 13, 14, and 15 illustrate the UMAP evidence visualisations. The number of Top K shown in these figures accurately corresponds to the evidence manually identified in the incident event knowledge base, as well as the interpretation of the similarity between the prompt and the events.

```
6    - VE (list or np.array): Embeddings of the
         Knowledge Base content.
7    Returns:
8    - torch.Tensor: Cosine similarity scores between
         input and Knowledge Base embeddings.
9    """
```

Figure 14 provides an example where all events are considered as evidence. The knowledge base for this scenario contains events where the attribute "Level" is set to "Warning". The ground truth data in Table X specifies a total of 25 pieces of evidence (Unauthorised Access Scenario), based on the user's input: *"Identify all events where the Level is 'Warning'"*. However, Figure 14, generated using UMAP, displays all 25 data points clustered together, indicating that both neighbourhood and context similarity are present. Despite this, the cosine similarity colour in the heat parameter for some data points appears colder, with a cosine similarity score of ≤ 0.54. This discrepancy suggests that, while all events share the same "Level" attribute, other features or contextual factors within the events (such as timestamps, event descriptions, or sources) contribute to their embedding representations, influencing their overall similarity. The colder colours indicate that, despite sharing the same "Level", these events may differ in other attributes or in how they relate to the query's context, particularly with the presence of the internal DFIR context-specific agent and its internally assigned role.

However, in this metric, we focus solely on evidence retrieval and identification, which has been successfully demonstrated, as all 25 pieces of evidence are clustered together. Another aspect to consider in the visualisation of Figure 14 is the size of Dimension 2, which ranges from 0 to 3.2, a range that is reduced when compared, for instance, to Figure 15. If increased, the data will appear closer together.

Table X contains the number of ground truth evidence (Incident Events - K) for each cyber incident, as determined by the user, the instructions prompted by the DFIR expert, and the Top K evidence generated by GenDFIR:

TABLE X: Top K Evidence within the Knowledge Base (Incident Events)

| Cyber Incident | Criteria of Evidence (Evidence Extraction Prompt) | Incident Events - K | Evidence - Top K |
|---|---|---|---|
| SYN Flood | Identify all Events with Level: Critical. | 30 | 05 |
| Rhino Hunt (Inspired from [61]) | Rhino. | 08 | 04 |
| Phishing Email - 1 | Identify all Events that appear to be phishing. | 15 | 06 |
| Phishing Email - 2 | Identify all Events that appear to be phishing. | 20 | 07 |
| Unauthorised Access | Identify all Events with Level: Warning. | 25 | 25 |
| DNS Spoof | Identify all Events with Level: Error. | 23 | 06 |

### C. GenDFIR Overall Performance

The overall performance of GenDFIR is evaluated by calculating the aggregate accuracy across each metric, as presented in Table XI:

| Metric | Rate |
|---|---|
| Accuracy | 95.52%. |
| Relevance | 94.51% |
| EM | 100% |
| Top-K | 100% |
| **Overall** | **97.51%** |

TABLE XI: Overall Performance

The results demonstrate a high performance rate, suggesting that, at this stage of development, the framework shows promise in terms of reliability and precision for DFIR investigations and Timeline Analysis. It is important to note, however, that these assessments were conducted under specific configurations and experimental conditions, which may affect their generalisability to other contexts.

### D. Discussion

According to our experiment, results, and the overall performance of the framework, and as previously discussed, GenDFIR should not be viewed as a universal solution for all DFIR tasks. Instead, it is designed as a foundational framework to address specific challenges faced by DFIR analysts or experts, particularly in the analysis of artefacts.

The evaluation results, as illustrated in the previous graphs, show a success rate ranging from 90% to 100%, depending on the metrics used. It is important to emphasise that these results are based on simplified experimental scenarios and minimal configuration settings. Moreover, the primary purpose of the generated report is to support the investigation process by providing general contextual knowledge and cybersecurity information related to the incident. In practice, casual and precise DFIR Timeline Analysis reports differ significantly, as both play a crucial role in documenting the sequence of events in an incident. These reports typically focus on:

- *Event Chronology:* A detailed timeline of events, presenting a chronological order of activities related to the incident.
- *Evidence Correlation:* Analysis of how various pieces of evidence relate to the timeline.
- *Incident Overview:* A summary of the incident, including key findings and impacts.

Additionally, these reports are often tailored to different audiences [64]:

- *Technical Stakeholders:* Forensics experts and IT professionals require detailed, technical reports with precise timestamps, technical analysis, and evidence correlation. For example, a technical report may include granular timestamps of system log entries and detailed forensic data.
- *Non-Technical Stakeholders:* Reports for non-technical audiences, such as senior management or legal teams, present a streamlined timeline with key events highlighted. These reports focus on the overall sequence of events and impacts, formatted to be accessible and understandable to non-experts. An example is an Executive Summary followed by a simplified timeline of events.

In the context of GenDFIR, the report templates were generated by the LLM, guided by both human prompts and the internal DFIR agent's prompt. Furthermore, the generated reports were constrained by the token limits of the selected LLM and the chosen embedding model. Consequently, if both the model and the token output size were altered to incorporate a larger model, the Timeline Analysis report would likely be longer and exhibit differences in content. However, the reports were automatically condensed within the scope of our current configuration, which is limited to a maximum of 2000 tokens due to the LLM's token constraints.

Another key aspect that underscores the utility of GenDFIR is its capability for evidence identification. This framework can readily identify evidence based on suspicions, assumptions, or specific elements intended to be monitored during the investigation. As previously demonstrated, evidence retrieval is facilitated by prompting the framework with context-specific DFIR instructions or queries, allowing for targeted identification of relevant events.

## VI. LIMITATIONS AND ETHICAL ISSUES

### A. Limitations

Despite the success of bringing the proposed framework to life and successfully generating timeline analysis reports, this research encountered several limitations:

- **Novelty of Approach:** The application of LLMs to DFIR Timeline Analysis is a relatively novel approach. The limited existing research on automating Timeline Analysis means there are few established guidelines or benchmarks, which has necessitated the development of a new context and methodology for this task.
- **Data Volume and Variety:** The large volume and heterogeneity of data in cyber incident scenarios present significant challenges for data management and processing [25]. Moreover, adapting solutions to real-world cyber incidents and DFIR practices is particularly difficult, as each incident introduces unique challenges that require tailored approaches. Data can vary considerably from one incident to another, and the rules that trigger each incident can differ [65], further complicating the solution. In the case of GenDFIR, our experiments were conducted using synthetic and oversimplified cyber incident scenarios, excluding anti-forensics techniques.
- **Privacy and Data Security:** Handling DFIR cases often requires access to sensitive personal data. To mitigate the risk of exposing real personal information, we have used synthetic data derived from synthetic cyber incidents as an alternative.
- **Evaluation Methods:** Assessing the effectiveness and performance of our framework presents significant challenges, particularly in the context of DFIR practices. Current research on automated evaluation methods for LLMs and AI systems in domain-specific applications is still in its early stages, and most approaches rely heavily on manual evaluation. In our case, this necessitated the creation of custom DFIR context-based evaluation prompts to measure performance. However, even with the proposed metrics, we contend that they are insufficient. The DFIR field is inherently complex and non-deterministic, with results often influenced by a wide range of unpredictable factors, such as varying incident types, data quality, and contextual nuances. These variables complicate the ability to derive consistent, controlled outcomes, making automated evaluation in DFIR an ongoing challenge. Moreover, the dynamic and rapidly evolving nature of cyber incidents demands evaluation methods that can adapt to real-world scenarios' technical and contextual variables, presenting an ongoing challenge to develop robust, reliable assessment frameworks.

### B. Ethical Issues

Introducing GenAI into the DFIR field presents serious ethical issues that are critically important and require meticulous consideration, given the sensitive and private nature of the data, practices, and activities involved. The following are the most relevant and common concerns [66]:

- **Privacy and Confidentiality:** The use of LLMs to advance cyber incident timeline analysis can lead to significant privacy breaches. These technologies often require access to and process vast amounts of sensitive data, including personal, digital, financial, health, and other types of information, all of which must comply with security standards and frameworks. This increases the risk of exposing personal and confidential information without proper safeguards.
- **Accuracy and Efficiency:** Ensuring the accuracy and reliability of automated analyses and generated reports from the LLM-powered framework is critical. Inaccurate results could lead to flawed or irrelevant conclusions and decisions by investigators and DFIR experts, potentially affecting legal proceedings and justice.
- **Consent:** Obtaining proper consent from individuals whose data is used for automatic processing in real-life scenario adaptations by the framework is essential. Without explicit consent, using such data would directly violate their privacy rights.
- **Bias and Fairness:** One major concern with AI models is their tendency to introduce or perpetuate biases, potentially leading to unfair outcomes in cyber incident analysis. For instance, in GenDFIR, the aspect of context enrichment tied to the interpretation of the knowledge base events can introduce bias and compromise results, impacting the impartiality and credibility of investigations.
- **Automatically Detected Evidence:** In cases where the framework is employed to detect and retrieve evidence autonomously, improper internal processing or analysis could compromise its reliability and validity.

## VII. CONCLUSION AND FUTURE WORKS

In this work, we present a novel AI-based framework designed to fully automate Timeline Analysis for cyber incidents. The framework integrates Retrieval-Augmented Generation (RAG) and large language models (LLMs) to not only conduct timeline analysis and generate detailed incident reports but also provide valuable insights and responses to various DFIR-related queries.

The research began with a comprehensive exploration of timeline analysis, examining its characteristics, the aspects requiring automation, and its informational cycle. We then proposed a development approach for the framework, addressing theoretical data representation and structuring as well as technical requirements, such as hardware and system configurations, to enable the framework to operate locally.

The framework was subsequently tested on synthetic DFIR incidents due to privacy and consent restrictions.

The evaluation of results highlighted the significant potential of the framework's core approach. However, notable limitations were encountered during this phase. Standard metrics traditionally used to evaluate machine learning (ML) and AI systems, such as F1-score, recall, and precision, were not directly applicable to the GenDFIR framework. This is because the framework produces contextual reports rather than numerical data for classification tasks, underscoring the need for bespoke metrics and evaluation methods specifically designed for GenDFIR.

While the current implementation demonstrates reliability, future work should focus on refining the framework by fine-tuning the LLM with context-specific data, rather than relying solely on a zero-shot approach. This should involve expanding experimentation to real-world scenarios, while ensuring strict adherence to security standards and data protection regulations. Furthermore, the results should be used to support advanced training of the LLM, with appropriate legal consent. Additionally, employing multiple specialised autonomous agents to independently manage distinct tasks—such as timeline analysis, evidence correlation, root cause identification, and the prediction of potential future threats—could further improve the system's precision and persistence. Exploring the use of larger models in optimised environments may also underpin more complex tasks and generate more detailed and accurate results.

In the previous sections, we have thoroughly explored the research questions and their implications. In this final section, we discuss the relationship between LLMs and Timeline Analysis. At this stage, the readiness of timeline analysis for LLM integration is closely tied to advancements in LLM technology. Progress in LLMs directly influences the effectiveness of timeline analysis automation, particularly in the generation of accurate timelines. The reports generated by GenDFIR for timeline analysis are probabilistic predictions based on token sequences, where the LLM predicts the next words in context. At this stage, these outputs cannot be fixed or directly controlled by a DFIR expert. Instead, the expert can only refine and guide the results through prompt engineering. While current LLMs demonstrate notable capabilities, they have not yet achieved complete accuracy and remain inherently tied to probabilistic predictions and their full potential for timeline analysis will only be realised with continued technological progress.

In conclusion, the proposed framework represents a significant advancement in applying GenAI to digital forensics and incident response, with the aim of enhancing digital security in an environment where cyber threats are continually evolving.

## Footnotes

### Ethical Approval

This research was deemed as not requiring the University's Ethical Committee Approval as it doesn't fall under any of the cases requiring ethical approval.

### Funding

The APC and Open Access fees for this publication are funded by the University of Liverpool.

### Acknowledgement

We acknowledge that all data used in this case study are synthetic used solely to model and visualise cyber incident scenarios while strictly adhering to privacy, confidentiality, consent, and copyright regulations.

All case studies, Knowledge Base (Incident Events - Ground Truth), GenDFIR Timeline Analysis reports, and evaluation results are available on https://github.com/GenDFIR/GenDFIR [62].

### Competing Interests

The authors declare that they have no known competing interests or personal relationships that could have appeared to influence the work reported in this paper.